\documentclass[11pt,preprint]{emulateapj}

\usepackage{amssymb}
\usepackage{amsmath}
\usepackage{wasysym}

\newcounter{minirefcount}
\newcommand{\mr}[2]{\refstepcounter{minirefcount}\label{#2}(\arabic{minirefcount}) #1}
\usepackage[title]{appendix}

\begin{document}

\title{A high resolution mid-infrared survey of water emission from protoplanetary disks}

\author{Colette Salyk}
\affil{Department of Physics and Astronomy, Vassar College, 124 Raymond Ave, Poughkeepsie, NY 12604, USA}

\author{John Lacy}
\affil{Department of Astronomy, University of Texas at Austin, Austin, TX 78712, USA}

\author{Matt Richter}
\affil{Physics Department, University of California at Davis, Davis, CA 95616, USA}

\author{Ke Zhang}
\affil{Department of Astronomy, University of Michigan, 311 West Hall, 1085 South University Avenue, Ann Arbor, MI 48109, USA }

\author{Klaus Pontoppidan}
\affil{Space Telescope Science Institute, 3700 San Martin Drive, Baltimore, MD 21218, USA}

\author{John S.\ Carr}
\affil{Naval Research Laboratory, Code 7213, Washington, DC 20375, USA}

\author{Joan R.\ Najita}
\affil{National Optical Astronomy Observatory, 950 N. Cherry Avenue, Tucson, AZ 85719, USA}

\author{Geoffrey A.\ Blake}
\affil{Division of Geological and Planetary Sciences, California Institute of Technology, MC 150-21, 1200 E California Blvd, Pasadena, CA 91125, USA}

\begin{abstract}
We present the largest survey of spectrally resolved mid-infrared water emission to date, with spectra for 11 disks obtained with the Michelle and TEXES spectrographs on Gemini North.   Water emission is detected in 6 of 8 disks around classical T Tauri stars.  Water emission is not detected in the transitional disks SR 24 N and SR 24 S, in spite of SR 24 S having pre-transitional disk properties like DoAr 44, which does show water emission \citep{Salyk15}.  With R$\sim$100,000, the TEXES water spectra have the highest spectral resolution possible at this time, and allow for detailed lineshape analysis.  We find that the mid-IR water emission lines are similar to the ``narrow component'' in CO rovibrational emission \citep{Banzatti15b}, consistent with disk radii of a few AU.   The emission lines are either single peaked, or consistent with a double peak.  Single-peaked emission lines cannot be produced with a Keplerian disk model, and may suggest that water participates in the disk winds proposed to explain single-peaked CO emission lines \citep{Bast11,Pontoppidan11}.  Double-peaked emission lines can be used to determine the radius at which the line emission luminosity drops off.  For HL Tau, the lower limit on this measured dropoff radius is consistent with the 13 AU dark ring \citep{Brogan15}.  We also report variable line/continuum ratios from the disks around DR Tau and RW Aur, which we attribute to continuum changes and line flux changes, respectively.  The reduction in RW Aur line flux corresponds with an observed dimming at visible wavelengths \citep{Rodriguez13}.
\end{abstract}

\section{Introduction}
Protoplanetary disk chemistry profoundly influences planetary properties, yet is difficult (perhaps impossible) to predict a priori for a given disk.  Observations of disk chemistry therefore provide the primary means to determine the chemical environments in which planets form.  The Atacama Large Millimeter Array is becoming the workhorse to study {\it outer} disk ($\gtrsim$ 10 AU) chemistry, but is not well suited for characterizing inner disks; inner disks generally contribute so little emission at millimeter wavelengths that large dynamic range is required to detect their contribution.  

Instead, inner disks (within a few AU) are best probed at infrared (IR) wavelengths, where molecules including CO, H$_2$O, OH, HCN, C$_2$H$_2$ and CH$_4$ \citep[e.g.][]{Najita03, Salyk08, Mandell12, Gibb13} have been detected.  With current telescopes, it is difficult to create images of molecular distributions; instead, indirect techniques are used.  With one method, molecules are observed across a range of wavelengths to measure how line fluxes vary with excitation energy, and the molecular distribution is reconstructed via comparison with a disk model \citep[e.g.][]{Zhang13}.  In a second method presented here, observed molecular Doppler shifts are converted to velocity, and Kepler's law is used to convert velocity to emitting radius \citep[e.g.][]{Salyk11b}.  This technique requires high spectral resolutions --- for example, to resolve gas at 1 AU around a solar-mass star with three resolution elements requires R$\sim$30,000.

Here we add to the study of physical and chemical conditions in inner disks by presenting the largest survey to date of spectrally resolved water emission at mid-IR wavelengths, obtained with the Michelle and TEXES spectrographs on the Gemini North telescope.  This work is complementary to the survey of near-IR water emission presented by \citet{Banzatti17}, and other near-IR water emission studies \citep{Najita00, Carr04, Thi05, Salyk08, Doppmann11, Mandell12}, as the lower excitation temperature of the mid-IR lines implies that they probe farther out in the disk.  Upper level energies of the detected mid-IR lines are in the range 3500--7000 K, while upper level energies for the detected near-IR lines are $>8000$ K \citep{Banzatti17}.  In addition, near-IR water observations of sources with low near-IR veiling are more difficult to interpret due to absorption from water-rich stellar photospheres \citep{Banzatti17}. This larger sample builds on the work of \citet{Pontoppidan10a} who detected mid-IR emission lines from two T Tauri stars, RNO 90 and AS 205 N, with the VISIR spectrograph on the VLT. However, with R$\sim$100,000, the TEXES spectrograph provides the highest current spectral resolution for mid-IR emission lines, allowing one to probe the kinematics of the water emission at radii of $\sim$a few AU.  Three of our spectra (for RW Aur, DoAr 44 and DR Tau) have been previously published in \citet{Knez07}, \citet{Salyk15} and \citet{Najita18}.

We begin by describing the observations and data reduction procedures in Section \ref{sec:observations}.  We then discuss the key findings of our study, including the prevalence of water in classical vs.\ transitional disks (\ref{sec:prevalence}), as well as the properties (\ref{sec:properties}) and location (\ref{sec:location}) of the observed water vapor.  We also discuss the emission line shapes and what this may imply about disk winds (\ref{sec:winds}) and line luminosities (\ref{sec:snow_lines}), and discuss variable emission from RW Aur (\ref{sec:variability_rwaur}) and DR Tau (\ref{sec:variability_drtau}).  We end with a discussion of some implications of this work, and of future prospects in this field (\ref{sec:discussion}), and we summarize our conclusions (\ref{sec:conclusions}).

\section{Observations and Data Analysis}
\label{sec:observations}
\begin{figure*}
\epsscale{1.3}
\plotone{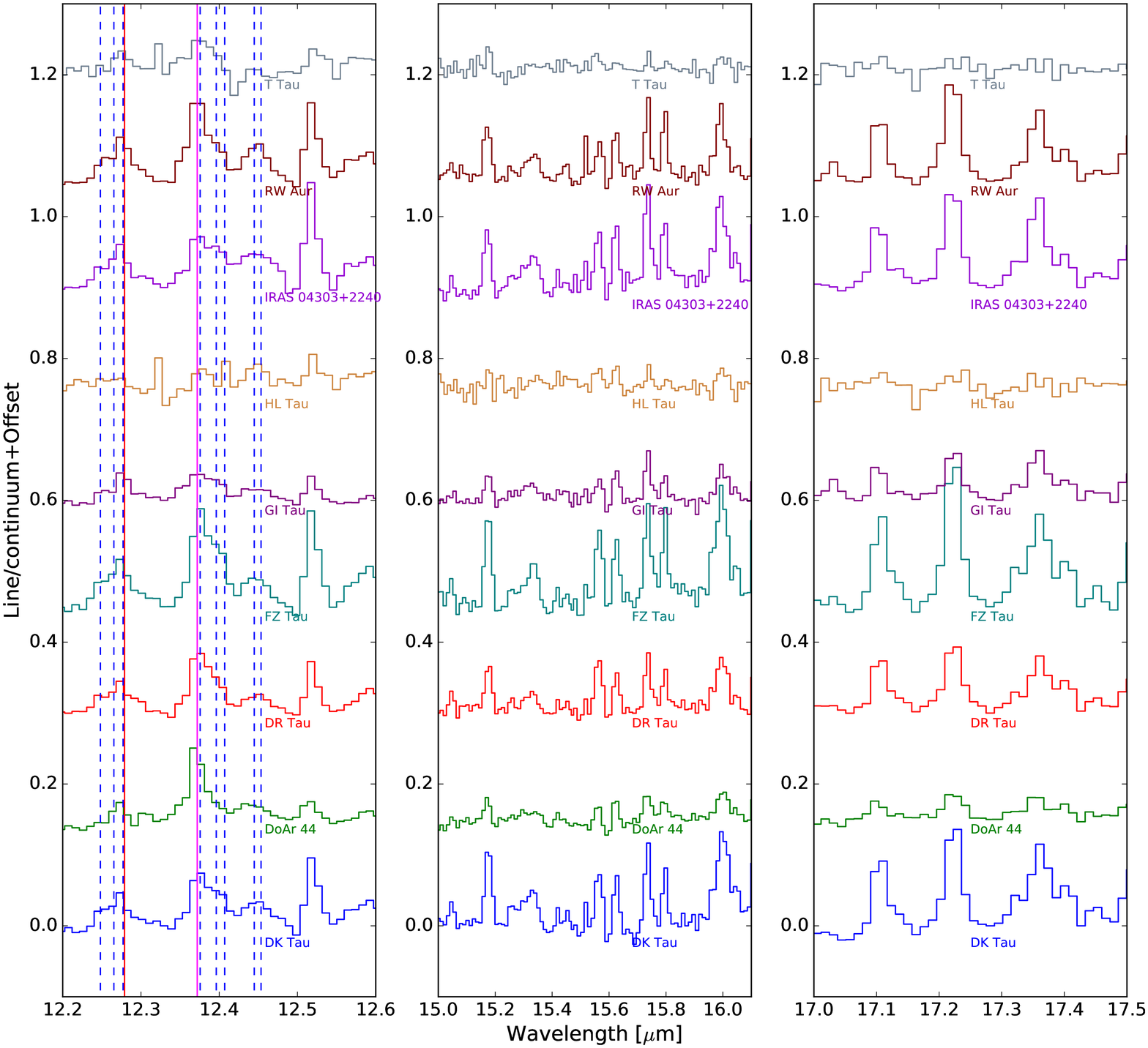}
\caption{Selected regions of continuum-subtracted and continuum-normalized Spitzer-IRS Short-High spectra for our sample sources, when available \citep{Pontoppidan10b,Carr11,Lebouteiller15}.  Blue vertical dashed lines in the left panel show locations of lines observed in this work. Solid red and magenta lines shows the locations of H$_2$ S(2) and H I 7--6, respectively.
\label{fig:Spitzer_closeup}}
\end{figure*}

A selection of T Tauri stars was observed with the Michelle \citep{Glasse97} and TEXES \citep{Lacy02} spectrographs on Gemini North between the years of 2011--2017 --- a complete observation log can be found in Table \ref{table:log}.  The spectra presented here represent a compilation of two complementary samples.  The first sample (DR Tau, FZ Tau, GI Tau, HL Tau, IRAS 04303+2240, RW Aur and T Tau N) is comprised of mid-IR bright T Tauri stars with mid-IR water emission observed with the Spitzer-IRS \citep{Pontoppidan10b,Carr11,Lebouteiller15}.  The second sample (DoAr 44, SR 24 N, SR 24S) is comprised of transitional disks (disks with low near-IR continuum fluxes), which were chosen to probe the effects of disk structure on the presence of water vapor in the inner disk.  The DoAr 44 data were previously presented in \citet{Salyk15}.  A compilation of source properties is shown in Table \ref{table:source_properties}.   Portions of the Spitzer-IRS spectra for our sample are shown in Figure \ref{fig:Spitzer_closeup}.  Most of the Spitzer spectra were reduced using routines developed by K. Pontoppidan \citep{Pontoppidan10b}.  Spectra of HL Tau and T Tau N are from the CASSIS database \citep{Lebouteiller15}, and only include the Short-High (SH) module. 

\begin{table*}
\caption{Observation Log}
\begin{tabular}{|l|l|l|l|l|l|l|l|l|}
\hline
Source & $\lambda_0$ [$\mu$m] &Airmass & UT Date & Exp Time [s] & Calibrator&Airmass&$\Delta v_\mathrm{tell}$ [km s$^{-1}]$\tablenote{Radial velocity shift between source and Earth: the sum of intrinsic and Earth-induced radial velocities.}&SNR\tablenote{Signal-to-noise ratio of divided (unbinned) spectrum in the wavelength range $12.252-12.261\,\mu$m, $12.38-12.392\,\mu$m or $12.46-12.48\,\mu$m, depending on the grating setting.}\\
\hline
\hline
TEXES&&&&&&&\\
\hline
\hline
DoAr 44 & 12.407\tablenote{Setting extends from $\sim$12.370--12.462 $\mu$m.} &1.43--1.70 & 2014 Aug 15,16&5700&Eunomia&1.52--1.63&22.3&10\\
DR Tau  & 12.407  & 1.10--1.30 &  2013 Nov 19&2461&Hygiea&1.29--1.31&16.5&14\\
FZ Tau  & 12.407   & 1.02--1.14  & 2013  Nov 19&2850&Hygiea&1.29--1.31&11.3&8\\
HL Tau  & 12.407    &1.07--1.11 & 2013  Nov 24&971&Davida&1.05--1.27&15.5&25\\
RW Aur A& 12.407  &1.03--1.09  & 2013 Nov 19&2072&Hygiea&1.29--1.31&5.0&8\\
RW Aur A& 12.407  &1.11--1.25, 1.25--1.33 & 2007 Oct 27,31 & 874,1036 & Flora & 1.12--1.13 & $-4.6$ & 5\\
SR 24 N & 12.407&1.41--1.42 & 2017 Mar 15&1036&Europa&1.20&$-33.6$&6\\
SR 24 S & 12.407& 1.41--1.42 & 2017 Mar 15&2072&Europa&1.20&$-34.8$&9\\
T Tau N  & 12.407 & 1.03&  2013 Nov 24&340&Davida&1.05--1.27&14.2&41\\
\hline
\hline
Michelle&&&&&&&\\
\hline
\hline

\hline
DK Tau                   &    12.378\tablenote{12.356--12.416 $\mu$m}     &1.11--1.66& 2011 Nov 11&2592&Eunomia&1.12--1.99&5.5&7\\ 
\hline
 DR Tau                  &    12.257\tablenote{12.233--12.296 $\mu$m}     &1.01--1.06&  2011 Jan 24&3024&Psyche&1.01--1.02&47.6&31\\ 
                               &    12.378                                                             &1.07--1.17&  2011 Jan 24&2160&Psyche&1.00--1.01&47.6&14\\ 
                               &    12.446\tablenote{12.424--12.482 $\mu$m}     &1.26--1.36&  2011 Jan 24&2160&Psyche&1.00&47.6&8\\ 
\hline
FZ Tau                    &    12.257                                                             &1.30--1.59& 2011 Jan 28&3024&Psyche&1.06--1.07&43.3&12\\ 
                               &    12.378                                                             &1.00--1.05 & 2011 Jan 30&4320&Psyche&1.02&43.0&22\\ 
\hline
GI Tau                     &   12.257                                                             &1.05--1.21& 2011 Jan 28&4320&Psyche&1.06--1.07&42.6&16\\   
\hline
IRAS 04303+2240  &   12.378                                                             &1.00--1.02&  2011 Nov 11&3024&Eunomia&1.09--1.12&8.2&8\\ 
\hline
RW Aur  A               &  12.257                                                              &1.02--1.03&  2011 Jan 25&3024&Psyche&1.08--1.09&37.3&59\\ 
                               &  12.378                                                              &1.14--1.31&  2011 Jan 25&3456&Psyche&1.11--1.13&37.3&47\\ 
                               &  12.446                                                              &1.35--1.53&  2011 Jan 25&2160&Psyche&1.04&37.6&18\\ 
\hline
\end{tabular}
\label{table:log}
\end{table*}

\begin{table*}
\caption{Source Properties}
\begin{tabular}{llllllllll}
Name & M$_\star$ & L$_\star$ & R$_\star$& d & i &log ($\dot{M}$)&$\log(L_\mathrm{acc})$\footnote{From provided references, when available, or calculated using $L_\mathrm{acc}=0.8\frac{GM_\star \dot{M}}{R_\star}$ from \citet{Gullbring98}}&v$_\mathrm{hel}$&Refs\\
 & [M$_\odot$] & [L$_\odot$] & [R$_\odot$] &[pc] & [$^\circ$] &[M$_\odot$ yr$^{-1}$]&[L$_\odot$] &[km s$^{-1}$]&\\
\hline
DK Tau & 0.52& 0.54 &1.53&129&$41\pm11$&$-8.3$/$-6.4$&$-1.2$/$0.72$&16.3&\ref{Gai16},\ref{Gai18},\ref{John02},\ref{Ngu12}\\
DoAr 44 &1.4 &1.3 &1.7&146&25&$-8.4$/$-8.0$ &$-0.88$/$-0.45$&$-5.9$&\ref{Andr09},\ref{Andr11},\ref{Gai16},\ref{Gai18},\ref{Mel03},\ref{Saly15}\\
DR Tau &0.4& 3.00&3.19& 196& 37 $\pm$ 3\tablenote{From 1.3mm continuum image.  Value used in our line shape models.}, 9\tablenote{From near-IR spectro-astrometry.}&$-6.68$/$-5.1$&$0.44$/$1.3$&21.1,27.6&\ref{Ard02},\ref{Gai16},\ref{Gai18},\ref{Isel09},\ref{John02},\ref{Ngu12},\ref{Pont11}\\  
FZ Tau &0.57 & 0.50&1.4 &130&38 $\pm$ 15&$-7.3$&$-1.3$&18.0&\ref{Banz15},\ref{Gah99},\ref{Gai16},\ref{Gai18},\ref{Pasc15},\ref{Rigl15},\ref{Saly13} \\
GI Tau & 0.60&0.33&1.09& 131&83& $-8.2$/$-6.9$&$-1.2$/$-0.62$&17.3&\ref{Gai16},\ref{Gai18},\ref{John02},\ref{Ngu12},\ref{Simo16}\\
HL Tau& 1.3& 3.0&3.0& 140& 46.72 $\pm$ 0.05 & $-8.8$/$-7.1$&$-1.3$/$0.41$&19&\ref{Brog15},\ref{Beck10}\\  
IRAS 04303+2240 & 0.43 & 1.49 &1.2& 148&...&$-6.1$&0.93&19.1&\ref{Andr13},\ref{Edwa06},\ref{Gai16},\ref{Gai18}\\  
RW Aur &1.48 & 1.60 &1.8&152&$45-60$ & $-6.7$/$-6.0$&$0.87$/$1.6$&15.9&\ref{Basr89},\ref{Dai15},\ref{Gah99},\ref{Gai16},\ref{Gai18},\ref{Hart95},\ref{Rigl15}\\  
SR 24 N &1.1&2.5&...&114& $121\pm17$ &$-6.9$&$0.03$&$-4.6$&\ref{Fern17},\ref{Gai16},\ref{Gai18},\ref{Natt06}\\
SR 24 S &1.5&2.5&...&114&  $70\pm5$&$-7.2$&$-0.23$&$-5.8$&\ref{Fern17},\ref{Gai16},\ref{Gai18},\ref{Natt06}\\
T Tau N & 2.11&  7.42 &3.30&144&$25\pm5$&$-7.5$&0.22&19.2&\ref{John02},\ref{Gai16},\ref{Gai18},\ref{Ngu12},\ref{Podi14}\\
\end{tabular}\\
\usecounter{minirefcount}{ 
\mr{ALMA partnership et al.\ 2015}{Brog15}
\mr{Andrews et al.\ 2009}{Andr09}
\mr{Andrews et al.\ 2011}{Andr11}
\mr{Andrews et al.\ 2013}{Andr13}
\mr{Ardila et al.\ 2002}{Ard02}
\mr{Banzatti \& Pontoppidan 2015}{Banz15}
\mr{Basri \& Bertout 1989}{Basr89}
\mr{Beck et al.\ 2010}{Beck10}
\mr{Dai et al.\ 2015}{Dai15}
\mr{Edwards et al.\ 2006}{Edwa06}
\mr{Fernandez-Lopez et al.\ 2017}{Fern17}
\mr{Gahm et al.\ 1999}{Gah99}
\mr{Gaia Collaboration et al.\ 2016}{Gai16}
\mr{Gaia Collaboration et al.\ 2018}{Gai18}
\mr{Hartigan et al.\ 1995}{Hart95}
\mr{Isella et al.\ 2009}{Isel09}
\mr{Johns-Krull \& Gafford 2002}{John02}
\mr{Melo et al.\ 2003}{Mel03}
\mr{Natta et al.\ 2006}{Natt06}
\mr{Nguyen et al.\ 2012}{Ngu12}
\mr{Pascucci et al.\ 2015}{Pasc15}
\mr{Podio et al.\ 2014}{Podi14}
\mr{Pontoppidan et al.\ 2011}{Pont11}
\mr{Rigliaco et al.\ 2015}{Rigl15}
\mr{Salyk et al.\ 2013}{Saly13}
\mr{Salyk et al.\ 2015}{Saly15}
\mr{Simon et al.\ 2016}{Simo16}
}
\label{table:source_properties}
\end{table*}

TEXES spectra were observed in the high-resolution cross-dispersed mode with R$\sim$100,000 (3$\,$km$\,$s$^{-1}$).  
Michelle spectra were observed in the high resolution mode, with R$\sim$25,000 (12.5 km s$^{-1}$).  
Spectra from both instruments were obtained in a standard manner for mid-IR spectroscopy.  To remove strong mid-IR sky background emission, the telescope nodded along the slit, and nod pairs were subtracted. To correct for telluric absorption, science observations were interspersed with observations of bright asteroids.  
 
TEXES spectra were reduced with the standard TEXES data reduction pipeline \citep{Lacy02}, with some modifications.  Nod pair differences were flat-fielded by dividing by the difference between a dome-temperature black-body and the sky emission.  Traces were corrected for curvature and spectra were extracted with optimal extraction.  Wavelength calibration was performed with custom-built atmospheric models, resulting in uncertainties of $\sim$ a few km s$^{-1}$.  Stellar spectra were then divided by asteroid spectra with correction for small differences in airmass.  This procedure resulted in good removal of atmospheric lines, but in some cases (notably HL Tau and T Tau N) left some tilt or curvature to the echelon orders.  This was removed by dividing each order by the median order spectrum.  
 
Michelle spectra were reduced with custom Python routines built by the authors.    Traces were corrected for curvature, and spectra were summed using optimal extraction.  Spectra were wavelength calibrated with telluric lines compared with Mauna Kea transmission models created by Gemini Observatory using the ATRAN modeling codes \citep{Lord92}.  The asteroid spectra were divided into the science spectra, after adjustment for small differences in airmass using Beer's law.   More details of the telluric division process for both TEXES and Michelle data can be found in Appendix \ref{Appendix}.

\begin{figure*}
\epsscale{1.3}
\plotone{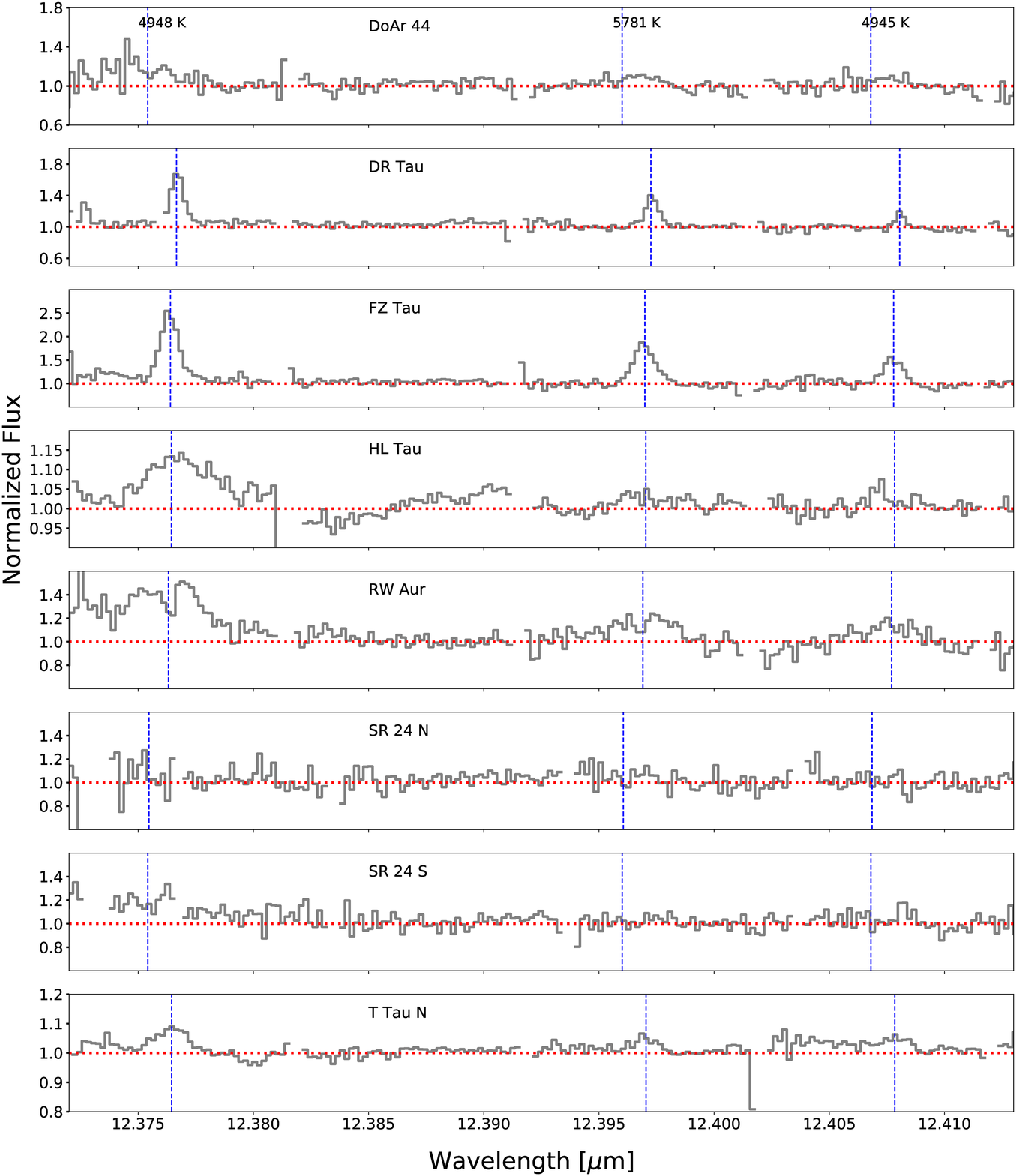}
\caption{Spectra obtained with TEXES spectrograph, binned to a resolution of 5 km s$^{-1}$.  Wavelengths are corrected for Earth-induced Doppler shifts.  Vertical dashed lines mark predicted locations of strong water lines (adjusted for heliocentric source velocities).  Horizontal dotted lines mark the continuum level.}
\label{fig:texes_all}
\end{figure*}

Observed spectra from Michelle and TEXES are shown in Figures \ref{fig:michelle_all} and \ref{fig:texes_all}, respectively.  In Figure \ref{fig:texes_all}, TEXES data are binned to a resolution of 5 km s$^{-1}$ to increase the signal-to-noise ratio (SNR).  Vertical lines mark expected locations for water emission lines, as listed in Table \ref{table:line_properties}.  Spectra show heliocentric wavelengths, and line locations have been adjusted for intrinsic source radial velocity (see Table \ref{table:source_properties}).  We estimate the radial velocities of most disks using published optical or UV stellar spectroscopy \citep{Ardila02,Gahm99,Melo03,Nguyen12,Pascucci15}.  For DR Tau, two significantly different values exist in the literature \citep{Ardila02,Nguyen12}, and in our plots and analyses, we use their average.  For SR 24 N, SR 24 S, and HL Tau, the source velocities come from mm-wave CO emission \citep{Brogan15,Fernandez17}.  For IRAS 04303+2240, we determine the source velocity using archival CO rovibrational emission spectra \citep{Salyk11b}.

\begin{table}
\caption{Detected Lines}
\begin{tabular}{|llll|}
\hline
J$_{K_aK_c}$ & $\lambda_0$ \footnote{Theoretical line centers from HITRAN molecular database \citep{Rothman13}.}& E$_\mathrm{upper}$ & Spin  \\
(upper $\rightarrow$ lower) & [$\mu$m]& [K] & \\
\hline
$17_{5\ 13}\rightarrow16_{2\ 14}$ &12.2481&5795&para\\
$18 _{7\ 12}\rightarrow17_{4\ 13}$&12.2654&6954&ortho\\
$12 _{8\  5}\rightarrow11_{5\  6}$  &12.2772&4048&ortho\\
$16_{4\ 13}\rightarrow15_{1\ 14}$ &12.3757&4948&ortho\\
$17_{4\ 13}\rightarrow16_{3\ 14}$ &12.3962&5781&ortho\\
$16_{3\ 13}\rightarrow15_{2\ 14}$ &12.4070&4945&para\\
$11_{8\ 3}\rightarrow10_{5\ 6}$    &12.4448&3629&ortho\\
$13 _{7\  6}\rightarrow12_{4\  9}$ &12.4535&4213&ortho\\
\hline
\end{tabular}
\label{table:line_properties}
\end{table}

We find that all detected emission lines are centered at the systemic velocity, with the possible exception of T Tau N (offset of +3 km s$^{-1}$).  However, this offset is comparable to the wavelength calibration uncertainty.  Therefore, we believe that the detected emission originates from the protoplanetary disk for all sources.

\section{Results}
\subsection{Prevalence of detections in transitional disks and classical disks}
\label{sec:prevalence}

\begin{figure*}
\epsscale{1.3}
\plotone{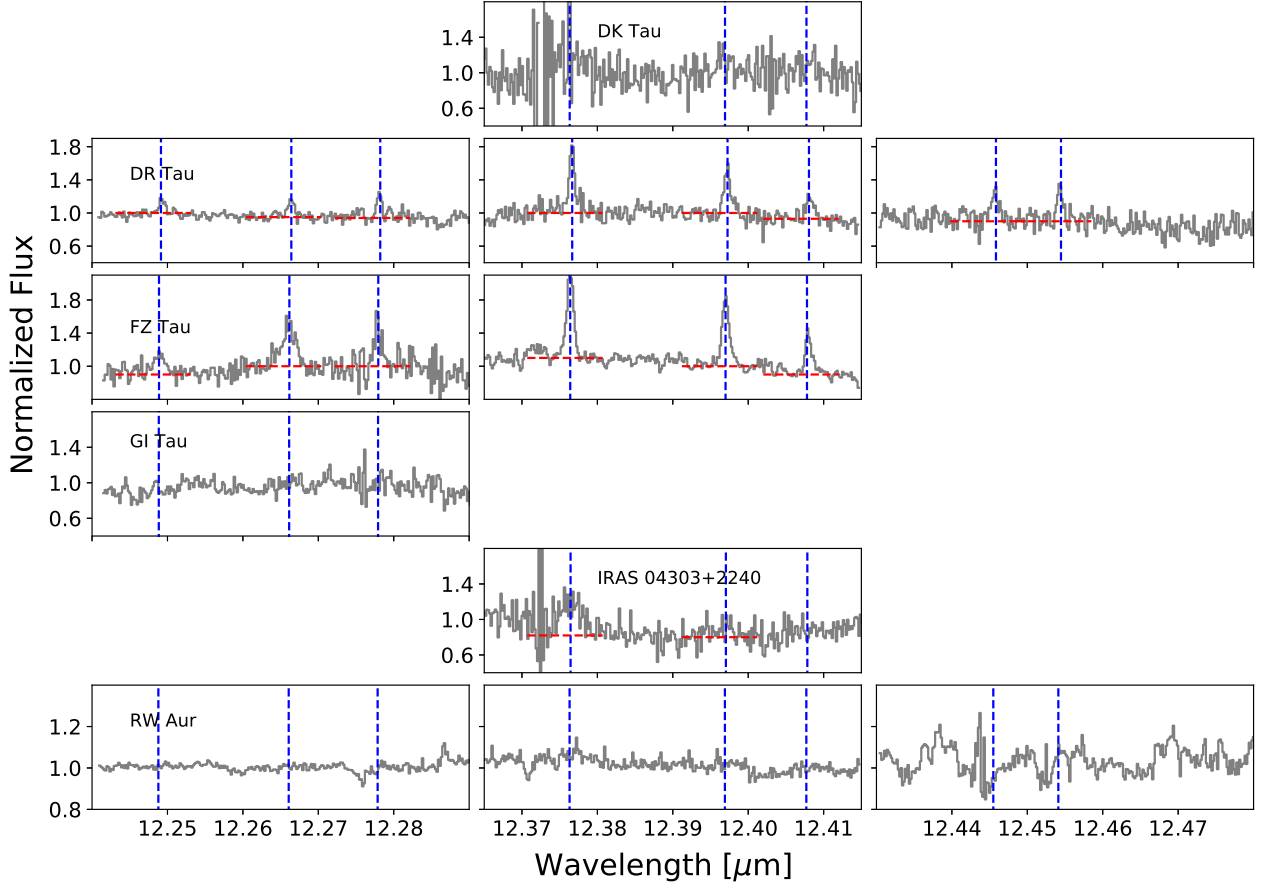}
\caption{Spectra obtained with Michelle spectrograph in three wavelength settings.  Wavelengths are corrected for Earth-induced Doppler shifts.  Vertical dashed lines mark predicted locations of strong water lines (adjusted for heliocentric source velocities).  Horizontal dashed red lines mark continuum levels used in flux calculations.}
\label{fig:michelle_all}
\end{figure*}

Deciding what to call a ``detection'' or ``non-detection'' is not always straightforward when uncertainties include difficult-to-quantify telluric correction errors, but we have made our best judgments based on our team's experience working with these types of datasets.  Within our chosen sample, water is detected in 6 out of 8 classical T Tauri disks and 1 out of 3 (pre-)transitional disks (which category includes DoAr 44, SR 24 N and SR 24 S).   For the classical T Tauri disk RW Aur, emission was detected in our TEXES data from 2013, and in prior data from 2007, but not in our Michelle data from 2011; this is discussed further in Section \ref{sec:variability_rwaur}.  The classical disks with non-detections, DK Tau and GI Tau, have the lowest continuum fluxes, and relatively weak mid-IR water emission fluxes \citep{Najita13}.  Figure \ref{fig:michelle_all} shows that DK Tau's non-detection is likely due to the low SNR of these data, while GI Tau's non-detection is likely due to its low line flux.   As described in Section \ref{sec:observations} and Figure \ref{fig:Spitzer_closeup}, all 8 of the classical disks, and one of the transitional disks (DoAr 44) were known water emitters based on Spitzer-IRS spectra.  Similarly, AS 205 and RNO 90, with VISIR-detected water emission, were known Spitzer-IRS water emitters with bright (7 Jy and 2.1 Jy, respectively) mid-infrared continuum flux levels \citep{Pontoppidan10a}.  Therefore, our results, combined with the works of \citet{Pontoppidan10a} and \citet{Najita18}, demonstrate that high-resolution follow up of molecular emission is feasible for bright, water-rich sources.

Amongst the transitional disks observed, SR 24 N and SR 24 S show no detectable emission, while DoAr 44 does show water emission.  SR 24 N is a close binary system (projected separation $0.08''-0.2''$, or $9-23$ AU; \citealp{Fernandez17}) that may have cleared much of the gas and dust from its interior disk.  SR 24 S shows a $\sim$32 AU gap in its millimeter continuum, although it retains a strong near-IR excess, indicative of the presence of small grains in the inner disk \citep{Andrews10}.   Previously, the SR 24 system had been observed with the Spitzer-IRS (Program 40145, PI: Houck), but only in the low-resolution mode, with which water vapor emission cannot be easily distinguished from the continuum.  In addition, the binary was unresolved.  DoAr 44 is a so-called ``pre-transitional'' disk, with an optically thick ring of dust close to the star \citep{Espaillat10}.  The non-detection of water in SR 24 N and SR 24 S is in keeping with an observed trend towards ``dry'' transitional inner disks \citep{Salyk15}.  Since the inner regions of transitional disks have reduced gas column densities and dust optical depths, the destruction of water vapor likely proceeds at a faster rate than its chemical production.   However, in the case of DoAr 44, the inner disk region must have high enough dust shielding and/or gas density to sustain a substantial water column, as is the case in classical T Tauri disks \citep{Salyk15}.

\subsection{Properties of 12 $\mu$m water emission}
\label{sec:properties}
Line fluxes and errors are shown in Table \ref{table:line_fluxes}. For TEXES spectra, we set the continuum level to 1 for the computation of all line fluxes.  For Michelle spectra, we define a local continuum for each line, as shown in Figure \ref{fig:michelle_all}.  Line wing velocities were chosen by eye, but are the same for all lines from the same source.  Based on tests with different choices of continuum levels and line wing limits, we estimate that statistical errors underestimate the integrated line flux errors by about a factor of 3, and we show these enhanced errors. In addition, the TEXES spectra include some order curvature that is not fully corrected by the standard star division, and that complicates the observation and characterization of some lines.  This is especially apparent in the spectra of HL Tau and RW Aur.  Finally, since our spectral data are not flux calibrated, line fluxes are computed assuming the 12 $\mu$m continuum flux values given in Table \ref{table:line_fluxes}.  Continuum flux variability in T Tauri stars of 20--50\% \citep{Kospal12} results in a corresponding uncertainty in line flux; this uncertainty is difficult to quantify without detailed photometric study of individual targets.

\begin{table*}
\caption{Line Fluxes}
\begin{tabular}{llllllllllll}
Source &Flux\tablenote{$10^{-15}$ erg cm$^{-2}$ s$^{-1}$, equivalent to $10^{-18}$ W m$^{-2}$.  Absolute fluxes assume the given $F_\mathrm{\nu,cont}$.  Errors do not include uncertainty in $F_\mathrm{\nu,cont}$.}&&&&&&&&F$_\mathrm{\nu,cont}$ \tablenote{Assumed continuum values from the references provided in the next column.} &Ref\tablenote{References for continuum measurements.} \\
&12.2481\tablenote{Line center in $\mu$m.}&12.2654&12.2772& 12.3757 & 12.3962&12.4070&12.4448&12.4535&[Jy]\\
&&&&&&&&&\\
\hline																											
DK Tau&&&&&&&&&0.86&2\\
\hline
DoAr 44&&&& $7.24 \pm 1.94$ &$2.76 \pm 1.93$ & $2.15 \pm 1.93$ &&&0.57&2\\
\hline
DR Tau &&&& $20.23 \pm 3.69$ & $10.92\pm3.68$ & $5.27\pm3.68$   &&&1.9&2\\ 
DR Tau\tablenote{Michelle fluxes.  Unless otherwise labeled, fluxes are from TEXES data.} &  $5.95  \pm 1.46$ & $7.79 \pm  1.76$ & $8.89  \pm 1.73$ &  $29.1  \pm 3.9$ &  $22.3 \pm 3.7$ & $11.3 \pm  4.2$ & $15.8 \pm 5.6$ & $13.5 \pm 4.7$&1.9&2\\
DR Tau\tablenote{From Banzatti et al.\ 2014 for three different epochs: 2011 Nov 27, 2011 Dec 2, and 2012 Jan 14.  If re-normalized to a continuum of 1.9 Jy, the fluxes for the three epochs would be 12.8, 15.3 and 11.4, respectively, for the 12.3962 $\mu$m line and 6.5, 8.8 and 6.3, respectively, for the 12.4070 $\mu$m line.} &&&&&$18.0\pm1.9$&$9.1\pm1.7$&&&2.67&1\\
DR Tau$^\mathrm{f}$ &&&&&$16.0\pm1.8$&$9.2\pm1.2$&&&1.99&1\\
DR Tau$^\mathrm{f}$ &&&&&$14.6\pm1.6$&$8.1\pm1.3$&&&2.44&1\\
\hline
FZ Tau  &&&& $28.26 \pm 4.37$ & $18.79\pm4.36$ & $10.39\pm4.36$ &&&1.0&2\\
FZ Tau$^\mathrm{e}$ &  $9.48  \pm  2.69$ & $21.29  \pm 2.82$ & $11.21  \pm  1.55$ &  $23.67 \pm 1.45$ & $20.68 \pm  1.38$ &  $11.10 \pm  1.19$&&&1.0&2\\
\hline
GI Tau  &&&&&&&&&0.86&2\\
\hline
HL  Tau &&&&  $98.0 \pm 10.7$    & $17.6\pm10.7$    &$19.1\pm10.7$    &&&11&3\\
\hline
IRAS 04303+2240$^\mathrm{e}$&&&& $32.7  \pm  10.3$&  $8.9  \pm 6.3$&&&&2.0&2\\
\hline
RW Aur &&&&  $56.9\pm5.1$     & $17.7\pm5.1$    &$8.34\pm5.1$     &&&1.5&2\\
\hline
SR 24 N &&&&&&&&&1.3&3\\
\hline
SR 24 S &&&&&&&&&1.5&3\\
\hline
T Tau N  &&&&  $79.4\pm13.1$    &  $37.4\pm13.1$   &$51.2\pm13.1$      &&&27&3\\
\hline
\end{tabular}
\label{table:line_fluxes}
(1) \citet{Banzatti14} (2) \citet{Pontoppidan10b} (3) \citet{Wright10}
\end{table*}

As shown in Table \ref{table:line_properties}, the emission lines at 12.3757 and 12.4070 $\mu$m, which are visible in the TEXES spectra, have similar excitation temperatures.  They also have similar Einstein A coefficients, but since the 12.3757 $\mu$m and 12.4070 $\mu$m lines are ortho and para water emission lines, respectively, they have a degeneracy ratio of 3.   For optically thin emission, therefore, the expected line ratio would be 3.  Given the error estimates in Table \ref{table:line_fluxes}, it is not possible to confirm whether the line ratios are consistent with optically thin or optically thick emission.  However, for the two disks with the best-determined line ratios (DR Tau and FZ Tau), the line ratios are consistent with being optically thin or marginally optically thick. 

Since a degeneracy exists between ortho/para ratio and optical depth, these data cannot be used to definitively quantify the ortho/para ratio.  However, since both low ortho/para and high optical depth would lower the 12.3757/12.4070 line ratio, the observed line ratios provide a lower limit to the ortho/para ratio.   All sources have ortho/para ratio lower limits consistent with 3, except for T Tau N.  For these sources, therefore, the ortho/para ratio is consistent with models of inner disk chemistry, in which water is produced rapidly via gas phase formation \citep[e.g.][]{Bergin07,Najita11}.

\begin{figure*}
\epsscale{1.}
\plotone{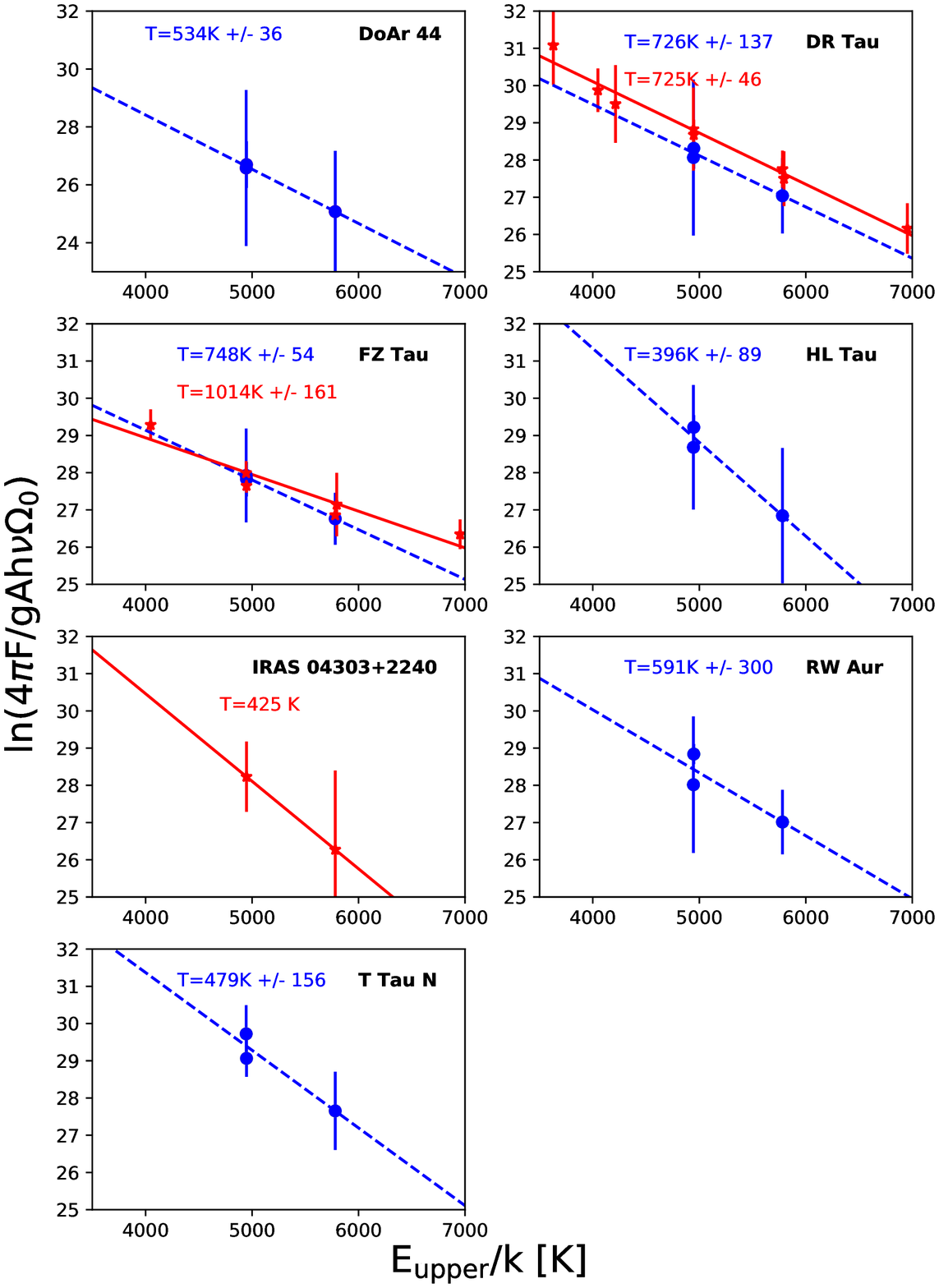}
\caption{Rotation diagrams and best-fit optically thin models.  Blue circles and dashed lines correspond to TEXES data, while red stars and solid lines correspond to Michelle data.  Model parameters are given in Table \ref{table:rotation_diagrams}.}
\label{fig:rotation_diagrams}
\end{figure*}

Line fluxes are displayed in the form of a rotation diagram in Figure \ref{fig:rotation_diagrams}.  For comparison with \citet{Pontoppidan10a}, we utilize the same y-axis scale, with values given in cgs units.  We assume $\Omega_0=\pi (1\ \mathrm{AU})^2/d^2$, where $d$ is the distance to the star, choosing a radius of 1 AU based on typical emitting radii for LTE models of mid-IR water emission \citep{Salyk11a,Carr11}.  When lines are optically thin, these rotation diagrams are linear, with a slope equal to $-1/T$ and intercept equal to $\ln (N_\mathrm{fixed-\Omega}/q(T))$ where $q(T)$ is the partition function (which is a function of temperature, $T$) and $N_\mathrm{fixed-\Omega}$ is the column density corresponding to the assumed $\Omega_0$.  Best-fit optically thin model parameters are shown in Table \ref{table:rotation_diagrams}.   Errors on temperature and column density account only for the scatter in the data and not the line flux error bars; therefore, the true uncertainty may be larger.  We also include optically thin fits to the VISIR-observed line fluxes in \citet{Pontoppidan10a}.  Derived rotational temperatures are in the range of $\sim$400--1000 K, consistent with the inner disk region.    

\begin{table*}
\caption{Model fits}
\label{table:rotation_diagrams}
\begin{tabular}{llllllll}
&High-Res&&&Spitzer\\ 
Source & T  [K] & $\log$ N$_\mathrm{fixed-\Omega}$\footnotemark[1] [cm$^{-2}$] & $\log$ $R_\mathrm{fixed-N}$\footnotemark[2]  [AU]& T [K] &$\log$ N [cm$^{-2}$] & $\log$ R\footnotemark[3][AU] &Spitzer Ref \\
DoAr 44 & $534\pm36$ &$18.23\pm0.22$&$0.12\pm0.11$&450&$18.3$&-0.036&2\\
DR Tau (T) & $726\pm137$&$18.04\pm0.37$&$0.025\pm0.19$&500&$18.6$&0.18&2\\
DR Tau (M) & $725\pm46$&$18.30\pm0.17$&$0.15\pm0.09$\\
FZ Tau (T) & $748\pm54$&$17.83\pm0.18$&$-0.09\pm0.09$&550&$19.0$&$-0.15$&3\\ 
FZ Tau (M) & $1014\pm161$&$17.36\pm0.27$&$-0.32\pm0.14$&\\
HL Tau  &$396\pm89$ &$20.41\pm0.60$&$1.2\pm0.3$&750&$19.0$&$-0.097$&3\\ 
IRAS 04303+2240 & $425$&19.8&$0.9$&250&$20.9$&0.43&3\\     
RW Aur & $591\pm300$&$18.7\pm2.0$&$0.35\pm1.0$&600&$18.3$&0.20&1\\ 
T Tau N &$479\pm156$&$19.8\pm0.7$&$0.9\pm0.4$&700&$19.6$&0.0&3 \\  
\hline
AS 205 & $520\pm75$ & $19.36\pm0.54$ &$0.68\pm0.27$&300 &$20.6$&0.34 &2\\
RNO 90 & $595\pm160$ & $18.11\pm0.99$ & $0.06\pm0.50$& 450&$18.3$ &0.23&2\\
\end{tabular}
\footnotetext[1]{Not a true column density.  The true column density, $N=N_\mathrm{fixed-\Omega} \frac{\Omega_0}{\Omega}$ where $\Omega_0=\pi (1\ \mathrm{AU})^2/d^2$ and $\Omega$ is the true emitting solid angle.  For $N_\mathrm{fixed-\Omega}\gtrsim10^{19}$ cm$^{-2}$, $N$ must be $<N_\mathrm{fixed-\Omega}$ for lines to remain optically thin.}
\footnotetext[2]{Emitting radius assuming $N=10^{18}$ cm$^{-2}$ and Area $=\pi R^2$.}
\footnotetext[3]{Assuming distance given in Table \ref{table:source_properties}} \\
(1) \citet{Carr11} (2) \citet{Salyk11a} (3) This work.  
\end{table*}

Some caution must be advised in interpreting the derived $N$.  If the lines are optically thin, differences in $N_\mathrm{fixed-\Omega}$ cannot be distinguished from changes in $\Omega$ -- i.e, stronger lines may arise from a higher column OR a greater emitting area.   Therefore, it is important to note that the true column density is given by $N=N_\mathrm{fixed-\Omega} \frac{\Omega_0}{\Omega}$ where $\Omega$ is the true emitting solid angle.   In addition, this procedure can be internally inconsistent, as $N\gtrsim10^{19}$ cm$^{-2}$ implies that some of the emission lines are optically thick.  $N$ {\it must} be $<N_\mathrm{fixed-\Omega}$ for T Tau N and HL Tau if the lines are optically thin.  For these reasons, we also show in Table \ref{table:rotation_diagrams} a calculation of emitting radius, $R$, assuming a fixed column density of $10^{18}$ cm$^{-2}$.  This calculation assumes that all of the variation in line flux arises from differences in emitting area, rather than differences in column density.  For the assumed $N$, the emitting radii are in the 0.5--few AU region of the disk. 

In order to compare with prior analyses based on Spitzer-IRS data, we also show parameters for LTE slab models that best fit the Spitzer-IRS spectra.  When available, we use model fits from the literature.  For sources where model fits do not appear in the literature (FZ Tau, HL Tau, IRAS 04303+2240 and T Tau N) we fit the spectra using the procedure outlined in \citet{Salyk11a}.   While most of the Spitzer-IRS best-fit models have temperatures in the range of 300--600 K, for HL Tau and T Tau N the best-fit tempeatures are 750 K and 700 K, respectively.  For these two sources, only SH spectra were available \citep{Lebouteiller15}; since the water emission actually arises from a range of disk radii, a fit to the higher excitation SH lines is likely to result in a higher best-fit temperature than a fit to the full Spitzer-IRS spectral range.  With this knowledge, we can see that the high-resolution spectra tend to have higher (by $\sim100-200$ K) best-fit temperatures than do fits to the full Spitzer-IRS spectral range.  This suggests that the high-resolution spectra presented in this work, which include only a few high-excitation lines, probe a relatively small, hot disk region; in contrast, the full range of the Spitzer-IRS high-resolution module, which includes a much wider range of upper level energies, probes a larger disk region.

\subsection{Location of water}
\label{sec:location}
\begin{figure*}
\epsscale{1.}
\plotone{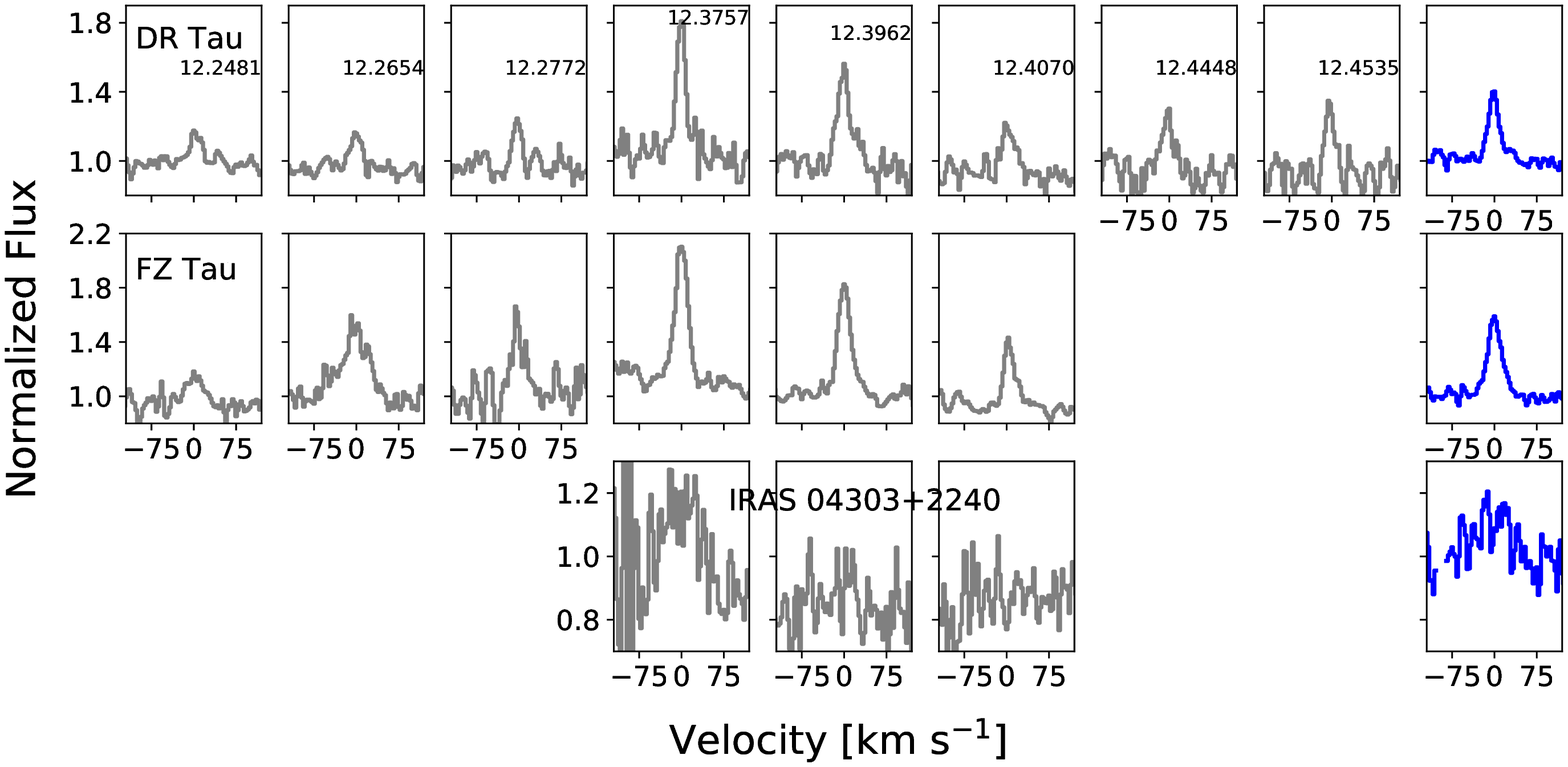}
\caption{Individual (gray) and composite (blue) water line profiles from Michelle spectra.  Top panel shows the theoretical central wavelength of each line, in $\mu$m. }
\label{fig:michelle_overlap}
\end{figure*}

\begin{figure*}
\epsscale{1.3}
\plotone{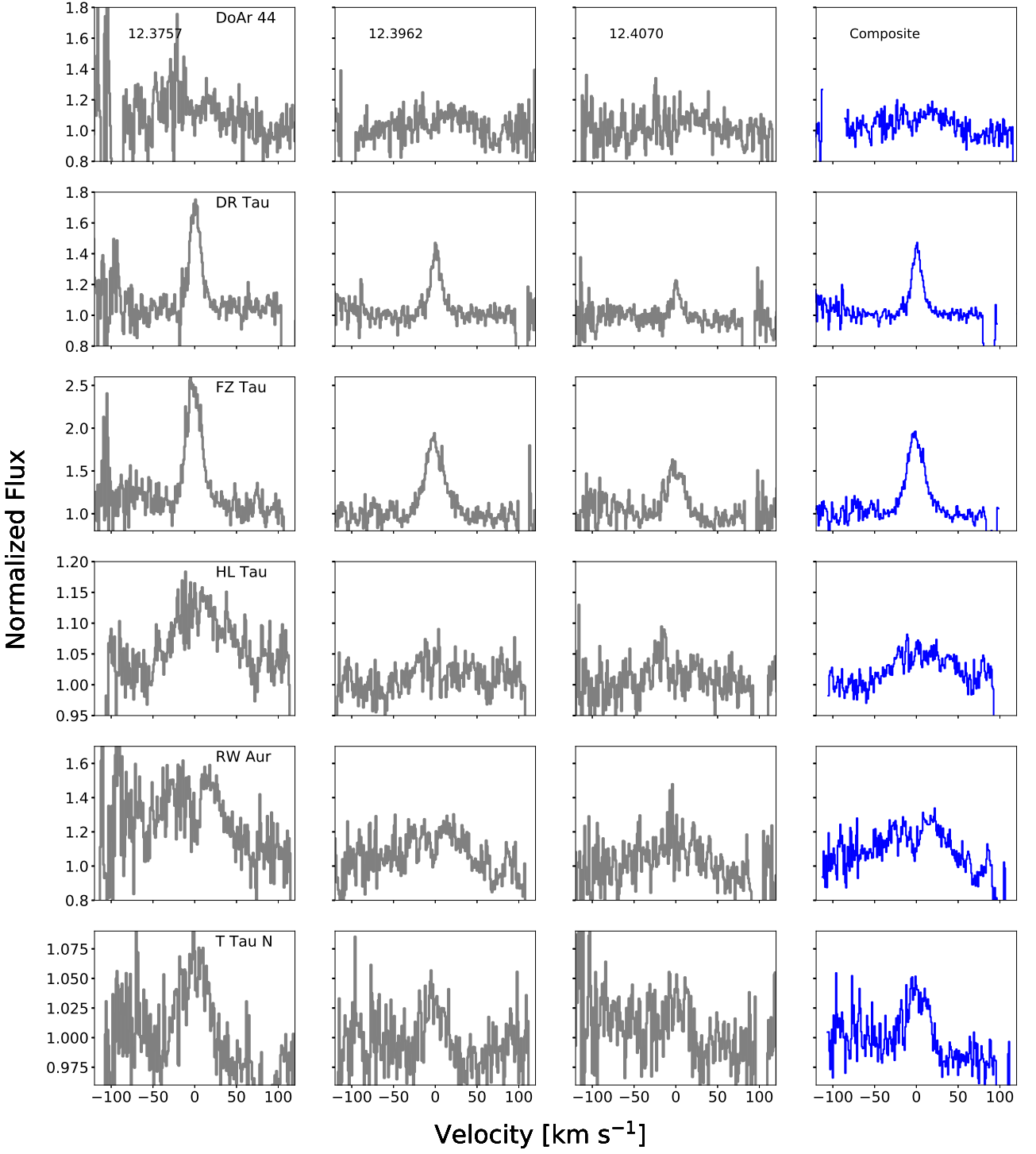}
\caption{Individual and composite water line profiles from TEXES spectra.}
\label{fig:texes_overlap}
\end{figure*}

To determine the location of the emitting water molecules more precisely, we analyze the shape of the emission lines, assuming they arise in a disk undergoing Keplerian rotation.   To analyze the emission line profiles with the highest possible SNR, we created line composites, shown in Figures \ref{fig:michelle_overlap} and \ref{fig:texes_overlap}, created by overlapping all detected emission lines, interpolating onto a common velocity grid, and averaging the spectra for each velocity value  (implicitly weighting by flux, thereby giving greater weight to stronger lines).  

\begin{figure*}
\epsscale{1.}
\plotone{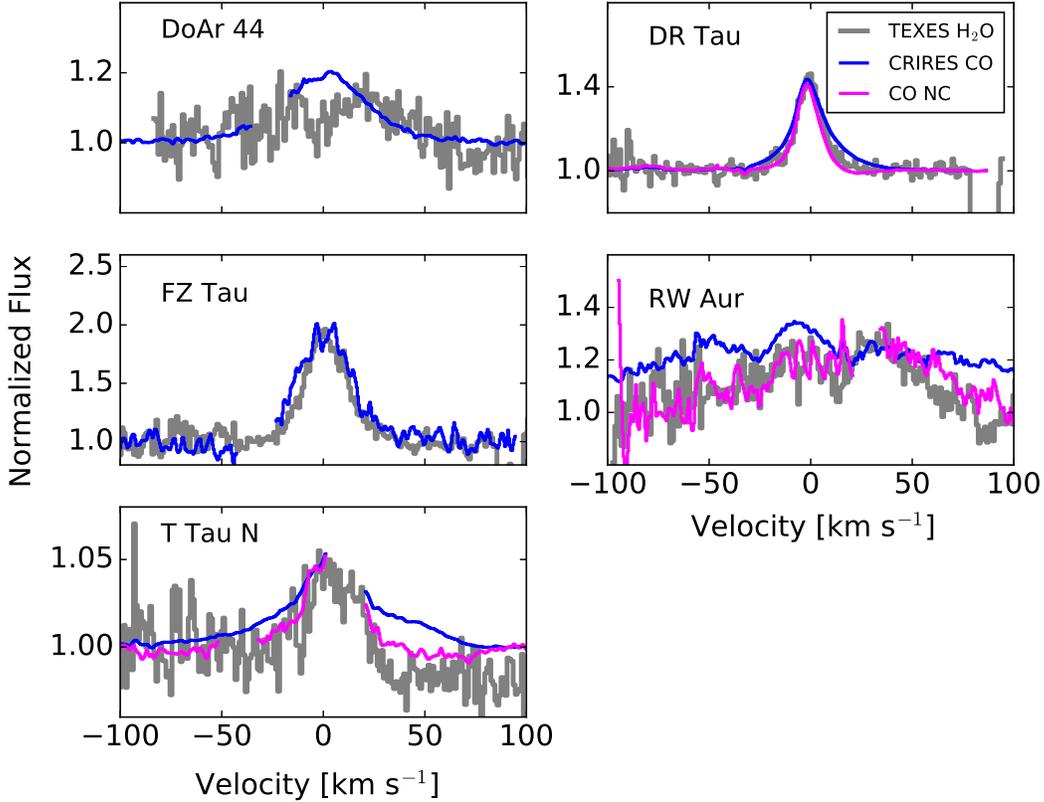}
\caption{Comparison of composite TEXES emission line profiles and CO fundamental emission line profiles obtained with CRIRES \citep{Banzatti15b}.  `NC' refers to the narrow component extracted from the CO profiles (see \citealt{Banzatti15b} for details).}
\label{fig:texes_crires_co}
\end{figure*}

There are 5 sources for which we have both TEXES water emission data and CRIRES CO rovibrational emission lines from \citet{Banzatti15b}.  We show a visual comparison of these emission lines in Figure \ref{fig:texes_crires_co}.  In all cases except for the transitional disk DoAr 44, the TEXES water emission lines are narrower than the CO emission lines, suggesting that the emission preferentially probes larger radii than the CO emission.  The CO  $v=1-0$ emission lines analyzed in that work probe slightly lower, but similar, upper level energies ($\sim$3000 K), as compared to the mid-IR water lines analyzed here.  \citet{Banzatti15b} fit the CO emission lines with either one or two (narrow and broad) components, with the broad component derived from $v=2-1$ emission lines (with upper level energies $\sim$6000 K).  For DR Tau, RW Aur and T Tau N, we find that the H$_2$O emission line profile is a closer match to the narrow component --- a correspondence predicted by \citet{Banzatti17}.   For DR Tau and T Tau N, the corresponding characteristic emitting radii reported by \citet{Banzatti15b} (derived from the Half Width at Half Maximum of the narrow component) are 0.5 and 1.1 AU, respectively (the RW Aur spectra were not included in their analysis).   DoAr 44 and FZ Tau are fit with a single component, with corresponding characteristic radii of 0.4 and 1.0 AU, respectively.  This comparison suggests that the water is emitted at $\sim$0.5--1.5 AU for these targets.  These results are consistent with the emitting radii derived from fitting the rotation diagrams with fixed column density, and with modeling of the velocity-unresolved Spitzer-IRS spectra \citep{Carr11, Salyk11a}.  

\begin{figure*}
\epsscale{1.}
\plotone{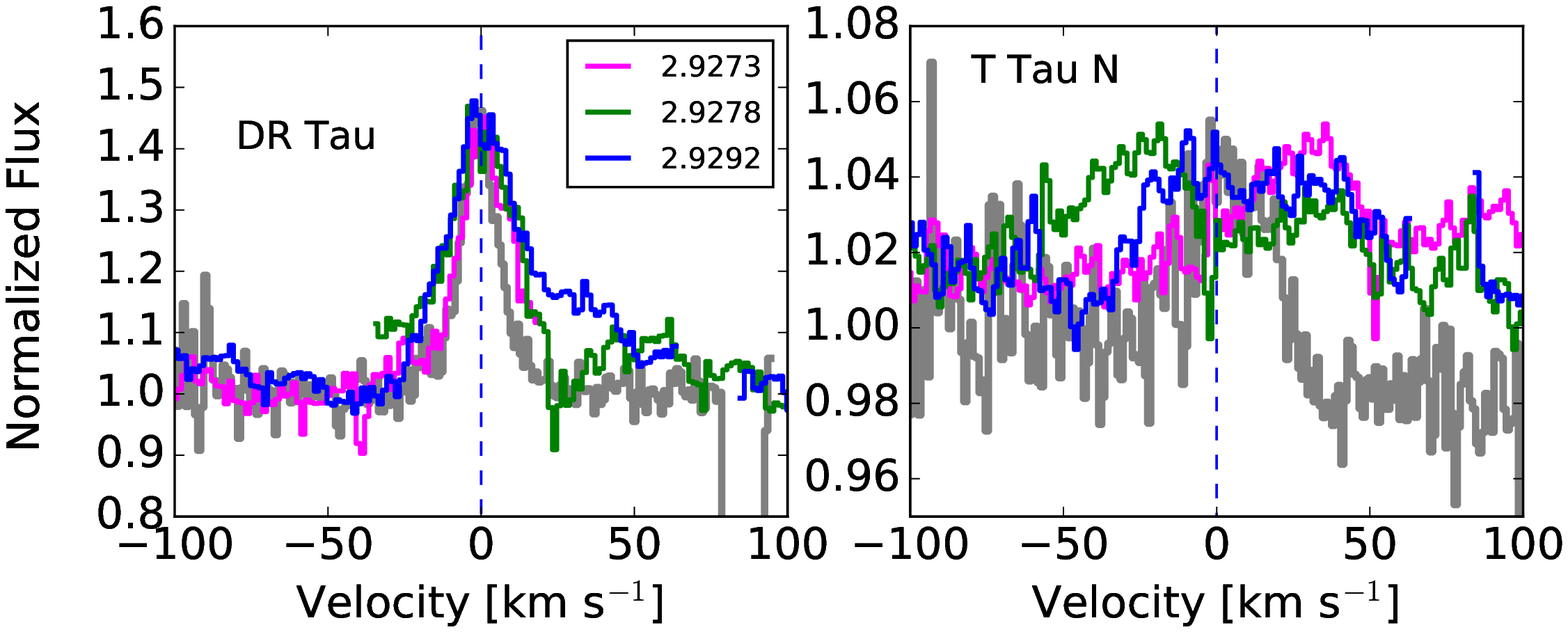}
\caption{Comparison of overlapped TEXES emission line profiles and rovibrational H$_2$O lines at 2.9273, 2.9278 and 2.9292 $\mu$m observed with CRIRES \citep{Banzatti17}.  }
\label{fig:texes_crires_h2o}
\end{figure*}

For DR Tau and T Tau N, 2.9 $\mu$m rovibrational water emission lines were also observed with CRIRES \citep{Banzatti17} and a comparison of CRIRES and TEXES lines is shown in Figure \ref{fig:texes_crires_h2o}.  For both sources, the TEXES emission lines are narrower than the CRIRES emission lines, although the 2.9 $\mu$m lines from T Tau N have structure that is not straightforward to interpret.  \citet{Banzatti17} note that the 2.9 $\mu$m lines correspond with the CO rovibrational broad component, which arises from the 0.04-0.3 AU region.  In contrast, the narrower 12 $\mu$m rotational water lines should arise from larger disk radii.

\begin{figure*}
\epsscale{1}
\plotone{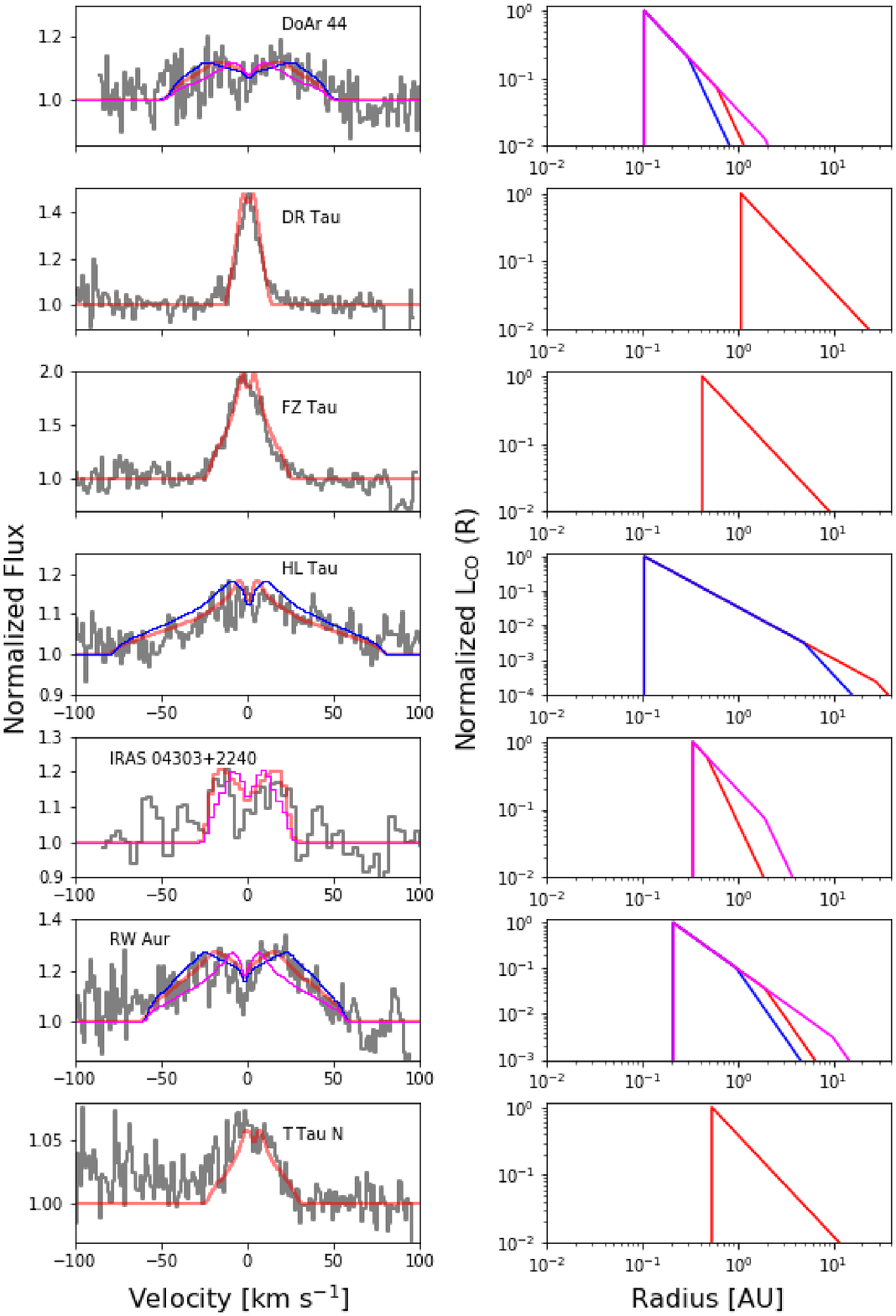}
\caption{Left: Observed line profiles with model fits.  Profiles are TEXES line composites except for HL Tau, for which we show only the 12.3757 $\mu$m line, and IRAS 04303+2240, for which we show a Michelle line composite.  Red curves show best-fit models with parameters shown in Table \ref{table:lineshape_model_fits}.  Blue and magenta curves show models with minimum and maximum values of $R_\mathrm{mid}$ (respectively) consistent with the observed lineshapes. Right: Luminosity profiles corresponding to each model. } 
\label{fig:texes_models}
\end{figure*}

We analyze the lineshapes quantitatively using the framework developed in \citet{Salyk11b} to model CO emission lines.  \citet{Salyk11b} utilize a simple parameterized model, in which the emission arises from a power-law luminosity profile, L(R), that drops off with radius.  Emission begins at a radius $R_\mathrm{in}$, extends to $R_\mathrm{mid}$ with luminosity $L(R)\propto R^{p}$ and to $R_\mathrm{out}$ with $L(R)\propto R^{q}$.  Following the analysis of \citet{Salyk11b}, we fixed $R_\mathrm{out}$ at 100 AU and $p=-1.5$, and we tested values of $q$ 
of $-3$,$-2.5$,$-2.0$, and $-1.5$.    Line wings are most sensitive to $R_\mathrm{in}$, while the location of double peaks are set by $R_\mathrm{mid}$.  Line fits are not very sensitive to $R_\mathrm{out}$, since L(R) decreases with radius.  Since the TEXES spectra have higher resolution, we analyze those lineshapes when available; for IRAS 04303+2240 we analyze the Michelle profile.  Best-fit models are shown in Figure \ref{fig:texes_models} and model parameters are shown in Table \ref{table:lineshape_model_fits}.  Since to our knowledge the inclination of IRAS 04303+2240 is unknown, we assume $i=45^\circ$ for our analysis.

For all targets, inner radii ($R_\mathrm{in}$) are between 0.1 and 1 AU --- i.e., the water arises in the terrestrial planet forming region of the disk.

\begin{table*}
\caption{Lineshape model fits}
\label{table:lineshape_model_fits}
\begin{tabular}{lllllll}
Source & R$_\mathrm{in}$&R$_\mathrm{mid}$&R$_\mathrm{mid}$ range&R$_\mathrm{out}$&p&q\\
&[AU] &[AU] &[AU]\\
\hline
DoAr 44 & 0.1 & 0.6 &0.3--2 &100 & $-1.5$ & $-3$  \\  
DR Tau & 1.0 &...&... &100&$-1.5$ &$-1.5$\\                
FZ Tau & 0.4 &...&...&100& $-1.5$  &$-1.5$\\                
HL Tau & 0.1 & 30 & $>5$&100& $-1.5$ &$-2$\\           
IRAS 04303+2240\tablenote{Not well characterized due to low SNR.} &0.3 &0.5 &0.5--2&100&$-1.5$&$-3$\\
RW Aur &  0.2 & 5 &1--10 &100 & $-1.$5 & $-3$\\     
T Tau N & 0.5&...&...&100&$-1.5$&$-1.5$\\                       
\end{tabular}
\end{table*}

\subsubsection{Single-peaked profiles}
\label{sec:winds}
The modeling highlights an interesting dichotomy in our sample:  DR Tau, FZ Tau, and T Tau N are all best fit with $q=-1.5$, implying that $R_\mathrm{mid}$ is undefined.  These line profiles are all single peaked, in contrast to expectations for Keplerian disks with moderate inclinations \citep{Horne86}, and in contrast to the remaining targets in the sample (DoAr 44, IRAS 04303+2240, HL Tau and RW Aur).   The sources with single-peaked line profiles have some of the lower inclinations in our sample, but the inclinations are still moderate, and one of the disks with a double-peaked profile (DoAr 44) has a low inclination.  Therefore, the line shape differences cannot be solely explained by viewing geometry.

Previous studies by \citet{Bast11} and \citet{Pontoppidan11} noted that DR Tau's combined single-peaked CO emission line shape and spatial profile could not be fit by a Keplerian disk model.    Both studies suggest that the velocity profile is modified by the presence of a slow molecular disk wind, as an angular-momentum conserving wind produces sub-Keplerian velocities \citep{Pontoppidan11}.  A combined disk+wind model, therefore, may produce broad line wings along with centrally-peaked emission.  \citet{Bast11} quantified the properties of the CO profiles with a ``peakiness'' parameter, $P_{10}$, defined as the full width velocity at 10\% of the line peak divided by the full width velocity at 90\% of the line peak.  They noted that $P_{10}\gtrsim 6$ is inconsistent with a Keplerian disk model with small spatial extent.  In contrast to the CO results, we find that the DR Tau and the other single-peaked {\it water} emission lines do not have large values of $P_{10}$, since the line profiles we observe are overall narrow, and do not have the wide base observed in CO. 

Nevertheless, in the context of our own modeling approach, a single peak implies that there is no defined drop-off in emission at $R_\mathrm{mid}$, or that this dropoff occurs far enough out in the disk that the double peaks are no longer resolved.  Our modeling shows that a dropoff radius of $R_{out}\gtrsim100$ AU would be required to produce double peaks small enough to remain unseen in our data (see Figure \ref{fig:texes_models}).  In contrast, \citet{Blevins16} find that the combined Spitzer-IRS and Herschel-PACS line {\it fluxes} from DR Tau and FZ Tau are best modeled with drops in water column density at $3.6\pm0.2$ and $3.3\pm0.2$ AU, respectively.  A model with a dropoff at 100 AU is therefore inconsistent with these results.  Note that the modeling approach we have adopted from \citet{Salyk11b} assumes that the disk luminosity decreases with radius. However, while single-peaked emission lines can be produced with a flat luminosity function, such models do not produce enough flux at high velocities to fit the line wings.  
Therefore, the velocity profile of the water emission lines, combined with the Spitzer and Herschel line fluxes, indicate sub-Keplerian motion.  One possible source of sub-Keplerian motion is a disk wind, as suggested by \citet{Pontoppidan11} and \citet{Bast11}.  

Disk winds may be expected to produce blueshifted emission lines, if the receding (redshifted) portion of the wind is hidden from view by the disk.  Table \ref{table:velocities} shows that observed water emission line velocities are consistent with both CO emission line velocities and stellar velocities (when known), to within a few km s$^{-1}$.  Therefore, if winds are the cause of the non-Keplerian lineshapes, the winds must have much lower radial velocities than are typically observed in jets and outflows --- a conclusion also reached by \citet{Pontoppidan11} and \citet{Bast11}.  Low radial velocities can result from low true velocities and/or low collimation.

\begin{table}
\caption{Heliocentric velocities of single-peaked emission lines. \label{table:velocities}}
\begin{tabular}{llllll|}
Source & V$_\mathrm{CO}$\tablenote{in km s$^{-1}$. Determined from CO rovibrational emission lines observed with NIRSPEC.}&V$_\star$ \tablenote{References in Table \ref{table:source_properties}} & V$_{\mathrm{H}_2\mathrm{O}}$\\
DR Tau & $25.6\pm0.5$ & $21.1,27.6$ & $25.6\pm3.0$ \\
FZ Tau & $15.9\pm0.5$ & $18.0$ & $17.1\pm3.0$ \\
T Tau N & $15.8\pm0.5$ &19.1&$17.4\pm3.0$ \\
\end{tabular}\\
\end{table}

The correspondence between water and CO line shapes is intriguing, since water vapor is easily photodissociated.   Therefore, if winds are the cause of the water line shapes, the wind must originate in a region with high enough densities of gas and/or dust to allow water production and/or shielding from UV radiation.  According to the models of \citet{Bethell09}, this occurs at gas column densities of $\sim10^{21}-10^{22}$ cm$^{-2}$.  

\subsubsection{Double-peaked profiles}
\label{sec:snow_lines}
The emission lines from DoAr 44, HL Tau, RW Aur and, perhaps, IRAS 04303+2240 (although the SNR for this source is low) show broader profiles, clearly showing or consistent with double peaks, implying a dropoff in emission strength at some disk radius.  In the context of our modeling approach, this dropoff is represented by $R_\mathrm{mid}$, as given in Table \ref{table:lineshape_model_fits}.  Given the SNR of the data, the value of $R_\mathrm{mid}$ is somewhat uncertain, and so we provide not only the best-fit value but also a range of values that appear consistent with the data.  The models with minimum and maximum values of $R_\mathrm{mid}$ are shown in Figure \ref{fig:texes_models}.  

A dropoff in emission line strength could in theory be indicative of a surface snow line --- i.e., the location on the disk surface where the water vapor freezes and becomes water ice.  However, the excitation temperatures of the observed 12 $\mu$m lines are $>$3500 K, while the water freeze-out temperature is $\sim$170 K.  Therefore, these lines should not effectively trace the freeze-out location, and the double-peaked lineshapes may simply reflect a steep disk temperature gradient with radius.  Nevertheless, if there is truncation of either the water vapor abundance or the overall gas density in the planet-forming region for any reason, it can be detected with these data.  We find, for example, that possible values of $R_\mathrm{mid}$ derived from our modeling include the 13 AU location of the first dark ring seen in images of the HL Tau disk \citep{Brogan15}, which has been attributed to a snow line \citep{Banzatti15a,Zhang15,Okuzumi16,Pinilla17} or an embedded planet \citep{Dipierro15,Dong15}.

With our small sample size, and the large error bars on $R_\mathrm{mid}$, it is difficult to make any firm conclusions about what these dropoff radii physically correspond to; however, this correspondence may be promising to pursue further with more targets and higher SNR data.  One way to distinguish an abundance or density dropoff from a steep temperature gradient would be to measure the peak spacing for individual lines with a range of excitation temperatures.  In addition, detailed comparisons of CO and H$_2$O line shapes in comparison with non-LTE disk models may enable a simultaneous determination of the inner disk temperature, density, and chemical structure.

\subsection{Variability in RW Aur}
\label{sec:variability_rwaur}
\begin{figure*}
\epsscale{1.}
\plotone{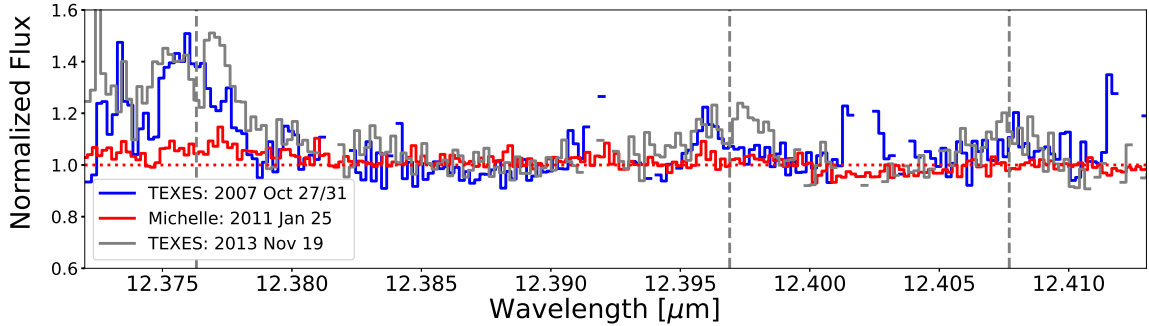}
\caption{Comparison of RW Aur spectra obtained in 2007, 2011 and 2013.}
\label{fig:rwaur_variability}
\end{figure*}

Michelle and TEXES spectra of RW Aur are markedly different, with TEXES data (from 2013 Nov 19, as well as archival data from  2007 Oct 27/31; \citealp{Knez07}) showing strong detections of water emission and Michelle data (from 2011 Jan 25) showing little to no emission --- see Figure \ref{fig:rwaur_variability}.  Acquisition images of the binary system confirm that RW Aur A was the observed target in both sets of observations.  

To distinguish between changes in line flux and changes in continuum flux, we can perform aperture photometry using Michelle acquisition images and the 12 $\mu$m fluxes given in Table \ref{table:line_fluxes}.  Unfortunately, the full set of photometric data (which includes DK Tau, FZ Tau and DR Tau) shows systematic offsets in $m_\mathrm{obs}-m_\mathrm{actual}$, indicating variability of these sources (or unknown systematic errors).   Therefore, photometry using each of these targets as references yields different continuum fluxes for RW Aur ---  $1.2\pm0.1$, $1.5\pm0.1$ or $6.1\pm0.5$ Jy for each of the three calibrators.  Since the Spitzer 12 $\mu$m flux and N-band flux ratio \citep{McCabe06} show that RW Aur A has a continuum flux of 1.4 Jy, the DK Tau and FZ Tau-based photometric values would imply that RW Aur A's continuum flux is ``normal'' during our observations. If the continuum flux has not changed significantly, the change in line/continuum ratio is a result of a reduction in emission line flux.

The Michelle observations coincide with a two magnitude V- and R-band dimming of RW Aur from September 2010 until March 2011 \citep{Rodriguez13}.  \citet{Rodriguez13} interpret the dimming as evidence of an eclipse of RW Aur by a tidal arm connecting RW Aur A and RW Aur B, matching the physical width of $0.27\pm0.05$ AU and distance of $180.5 \pm 28.9$ AU implied by the transit lightcurve shape.  

Variability in observed line/continuum ratios of the water emission lines can have a few explanations.  Firstly, the 12 $\mu$m continuum flux could have increased.  The lack of concurrent photometric data makes it difficult to rule out this possibility, but, as the analysis of our acquisition images shows, it is likely that the 12 $\mu$m continuum has not changed significantly.  If instead the line fluxes have changed, there are several possible explanations.    A physical change could have affected the temperature structure or degree of settling in the upper disk atmosphere, reducing the column of hot gas required to produce strong emission.  Or, a chemical change could have reduced the abundance of water in the upper atmosphere.   This was observed to occur in the variable EX Lupi \citep{Banzatti12}, interpreted as a destruction of water vapor via photodissociation, but was associated with an outburst, rather than dimming, event.  Or, the gas excitation level could have changed --- perhaps due to a change in accretion luminosity.

Given the tidal tail blocking hypothesis presented by \citet{Rodriguez13}, we consider whether the line emission could have a different physical location than the continuum emission and its emission could be physically blocked.  The 12 $\mu$m continuum corresponds to the peak of a $\sim$240 K blackbody.   \citet{Chiang97} predict a disk (dust) surface temperature of $\sim$240 K at $\sim$8.5 AU for a star with radius 1.8 R$_\odot$ and temperature 4730 K \citep{Rigliaco15}.  If the line emission arises from the $\sim$1 AU region, it is at least plausible that a 0.27 AU wide tidal arm may block a significant amount of the line emission while blocking less of the 12 $\mu$m continuum.   Given the moderate (45-60$^\circ$) inclination of the disk, this would require that the tidal arm extends significantly beyond the plane of the disk, as also suggested by \citet{Rodriguez13}.

\subsection{Variability in DR Tau}
\label{sec:variability_drtau}
\begin{figure*}
\epsscale{1.}
\plotone{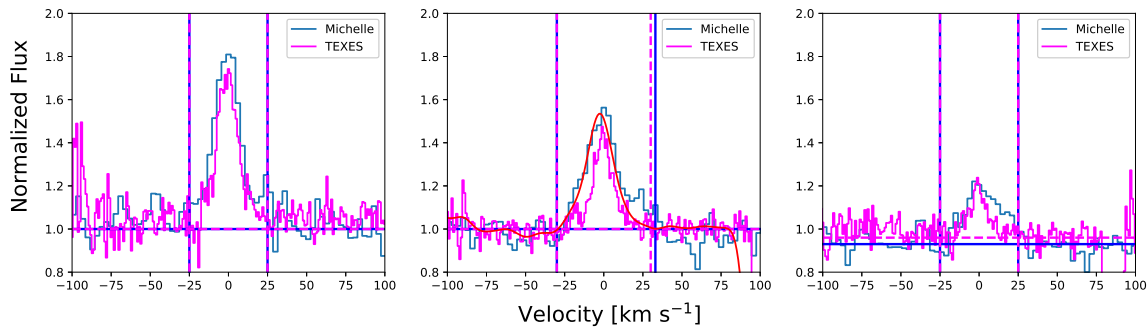}
\caption{Comparison of DR Tau spectra obtained with Michelle and TEXES in 2011 and 2013, respectively.  Vertical and horizontal solid blue and dashed magenta lines show the boundaries utilized for flux calculations.  In the middle panel, the red curve shows the TEXES emission line shifted by 2 km s$^{-1}$, multiplied by 1.7, and convolved with a Gaussian spectral response function with FWHM=12.5 km s$^{-1}$.}
\label{fig:drtau_comparison}
\end{figure*}

As shown in Figure \ref{fig:drtau_comparison}, Michelle and TEXES spectra of DR Tau show some differences, with Michelle line/continuum ratios being higher by about a factor of 2.  \citet{Banzatti14} investigated variability in the water spectra of DR Tau from Nov 2011 to Jan 2012, and noted $\sim$30\% changes in continuum-normalized line fluxes, which they attributed to changes in the N-band continuum.  Without photometric standards for both sets of observations, we cannot say whether the observed changes in line-to-continuum ratios are due to line flux changes or continuum changes.  Nevertheless, as described above, photometry using the Michelle acquisition images of DR Tau, DK Tau, and FZ Tau suggests that the DR Tau continuum flux is low (0.4--0.5 Jy) in the Michelle spectra.  This is quantitatively consistent with the factor of two change in line-to-continuum if the continuum level for the TEXES observations is also lower than the Spitzer-derived value by a factor of 2.  Although \citet{Banzatti14} report fluxes different from ours by several sigma, re-normalization to a continuum of 1.9 Jy would make the line fluxes consistent with our TEXES fluxes --- consistent with a picture in which the line fluxes are not changing, but the continuum is.  

The spectra also show a 1--2 km s$^{-1}$ velocity offset, which is within the expected uncertainty for wavelength calibration.  The Michelle lineshapes are also wider, but this is consistent with the Michelle Spectral Response Function.  The second panel in Figure \ref{fig:drtau_comparison} shows the TEXES emission line shifted by 2 km s$^{-1}$, multiplied by 1.7, and convolved with a Gaussian spectral response function with FWHM=12.5 km s$^{-1}$, demonstrating that the two sets of observations have similar intrinsic lineshapes.

\section{Discussion: Current and future potential for mid-IR water spectroscopy}
\label{sec:discussion}
With this work, we show that existing high-resolution mid-IR spectrographs on 10-meter class telescopes can successfully detect water vapor emission lines from disks around low mass stars.  In particular, we find detectable levels of emission from disks around 6 of 8 classical T Tauri stars in our sample (with the TEXES spectrograph detecting 6 of 6 observed), and from one of three transitional disks (DoAr 44).  The 12$\mu$m continuum for DoAr 44 is only 0.6 Jy, and there remain a significant number of potential targets with continuum levels greater than this value.  Therefore, the possibilities for this type of work are not yet exhausted with current facilities.

The TEXES instrument also provides the highest mid-IR resolution of any available instruments.  This unique capability allows us to investigate the gas kinematics, including any non-Keplerian motions (as may occur in a disk-launched wind), or to measure any depletions of water vapor in the disk surface.  It is clear from this work that obtaining a high SNR for the spectral data is crucial for constraining the locations and shapes of the line peaks. 

One limitation of this type of study is that it probes a relatively small range of excitation temperatures and optical depths.  Studies of the column density and temperature of water vapor throughout the disk are better suited to space-based instruments, which can probe a wide range of excitation conditions without being limited by the interference of Earth's atmosphere.  Therefore, this work remains complementary to studies performed with the Spitzer-IRS and future studies with the James Webb Space Telescope.  The ideal instrument for this type of work would be a high-resolution spectrograph on a space telescope, which may be a possibility for the Origins Space Telescope \citep{Pontoppidan18}.

Another limitation of the current generation of high resolution water spectroscopy is that the sources must have sufficiently strong mid-IR emission lines.  Even if the lines are detected, analysis of line shapes can be inhibited by low SNRs.  The availability of a mid-IR high resolution spectrograph on a 30-meter class telescope would instead make it possible to study virtually all molecule-rich disks in the nearest star forming regions with high SNRs.  Such a facility would also allow for the observation and analysis of less prominent lines, such as HCN and C$_2$H$_2$ \citep[e.g.][]{Najita18}.

\section{Conclusions}
\label{sec:conclusions}
In conclusion, we find:
\begin{itemize}
\item {\bf High resolution mid-IR water emission is readily detected around bright, water-rich classical T Tauri stars.}  Water is detected in 6 of 8 classical T Tauri stars in our sample, but water emission is detected in only one ``transitional'' disk, DoAr 44.  As DoAr 44 has a ring of optically thick material close to the star, it is distinct from other transitional disks, and has been categorized as a ``pre-transitional'' disk \citep{Espaillat10}.  Therefore, our observations are in keeping with a scenario in which the cleared inner regions of transitional disks can no longer support the presence of water vapor, due to a combination of reduced gas column densities and dust shielding \citep{Salyk15}.  However, if the inner disk region has high enough dust shielding and/or gas densities (as is true for both classical T Tauri disks and pre-transitional disks), it remains water rich.
\item {\bf Both excitation and line shape analyses demonstrate that the mid-IR water lines are emitted from the few AU region of the disk}, confirming results from smaller initial studies \citep{Knez07,Pontoppidan10a,Banzatti14,Salyk15}. While suggested by excitation studies with the Spitzer-IRS \citep{Carr11, Pontoppidan10b}, spectrally resolved observations confirm this result directly with gas kinematics.  Observations of mid-IR emission lines are complementary to near-IR emission studies, which probe closer to the star.  The mid-IR emission lines are similar to the ``narrow component'' seen in CO rovibrational emission \citep{Banzatti15b}.  
\item {\bf A subset of the sample shows single-peaked emission lines.  If disk winds explain single-peaked CO rovibrational lines \citep{Bast11, Pontoppidan11}, water may be participating in this same process.}  The observed single-peaked lines cannot be reconciled with Keplerian disk models, in which the line luminosity would decrease with radius.  An alternative explanation is that a sub-Keplerian wind ``fills-in'' the line profile, making it appear single-peaked.  If this is the case, the wind launching region must be dense enough and shielded enough to support the retention of water vapor.
\item {\bf A subset of the sample is consistent with double-peaked emission lines.}  For HL Tau, the lower limit of 5 AU on the dropoff radius (the location where the line luminosity begins to drop more steeply) is consistent with the measured 13 AU dark ring \citep{Brogan15} that may be attributed to a snow line \citep{Zhang15,Pinilla17} or an embedded planet \citep{Dipierro15}.  Higher SNR data would be necessary to confirm this result, and to determine the physical explanation for the profile shapes.
\item {\bf We find variable emission line strengths from the disk around RW Aur, which coincides with a visible two magnitude dimming event \citep{Rodriguez13}.}  These observations are consistent with the tidal arm blocking scenario suggested by \citet{Rodriguez13} if the tidal arm blocks the inner disk region producing the water emission, while only partially blocking the larger disk region producing the 12 $\mu$m continuum.
\item {\bf The possibilities for studying mid-IR water vapor emission with existing facilities are not yet exhausted.  Nevertheless, the availability of a high resolution mid-IR spectrograph on a 30-meter telescope would expand the potential sample to include most low mass stars in the nearest star forming regions.}  Observations of resolved line profiles with high SNRs have the potential to reveal detailed information about inner disk physics and chemistry.
\end{itemize}

\section{Acknowledgements}
The authors thank the anonymous reviewer for thoughtful suggestions that improved the manuscript.  The authors would like to thank Marie Lemoine-Busserolle and Rachel Mason for their support during the Michelle observations at Gemini North. C.S. would like to thank Andrea Banzatti for providing CRIRES data and for helpful discussions.  K.Z. acknowledges the support of NASA through Hubble Fellowship grant HST HF2-51401.001-A awarded by the Space Telescope Science Institute, which is operated by the Association of Universities for Research in Astronomy, Inc., for NASA, under contact NAS-26555. This work is based on observations obtained at the Gemini Observatory, which is operated by the Association of Universities for Research in Astronomy, Inc., under a cooperative agreement with the NSF on behalf of the Gemini partnership: the National Science Foundation (United States), the National Research Council (Canada), CONICYT (Chile), Ministerio de Ciencia, Tecnología e Innovación Productiva (Argentina), and Ministério da Ciência, Tecnologia e Inovação (Brazil).  This publication makes use of data products from the Wide-field Infrared Survey Explorer, which is a joint project of the University of California, Los Angeles, and the Jet Propulsion Laboratory/California Institute of Technology, funded by the National Aeronautics and Space Administration. This work has made use of data from the European Space Agency (ESA) mission Gaia (https://www.cosmos.esa.int/gaia), processed by the Gaia Data Processing and Analysis Consortium (DPAC, https://www.cosmos.esa.int/web/gaia/dpac/consortium). Funding for the DPAC has been provided by national institutions, in particular the institutions participating in the Gaia Multilateral Agreement.

\begin{appendices}

\section{Telluric Correction}
\label{Appendix}

\begin{figure*}
\epsscale{1.}
\plotone{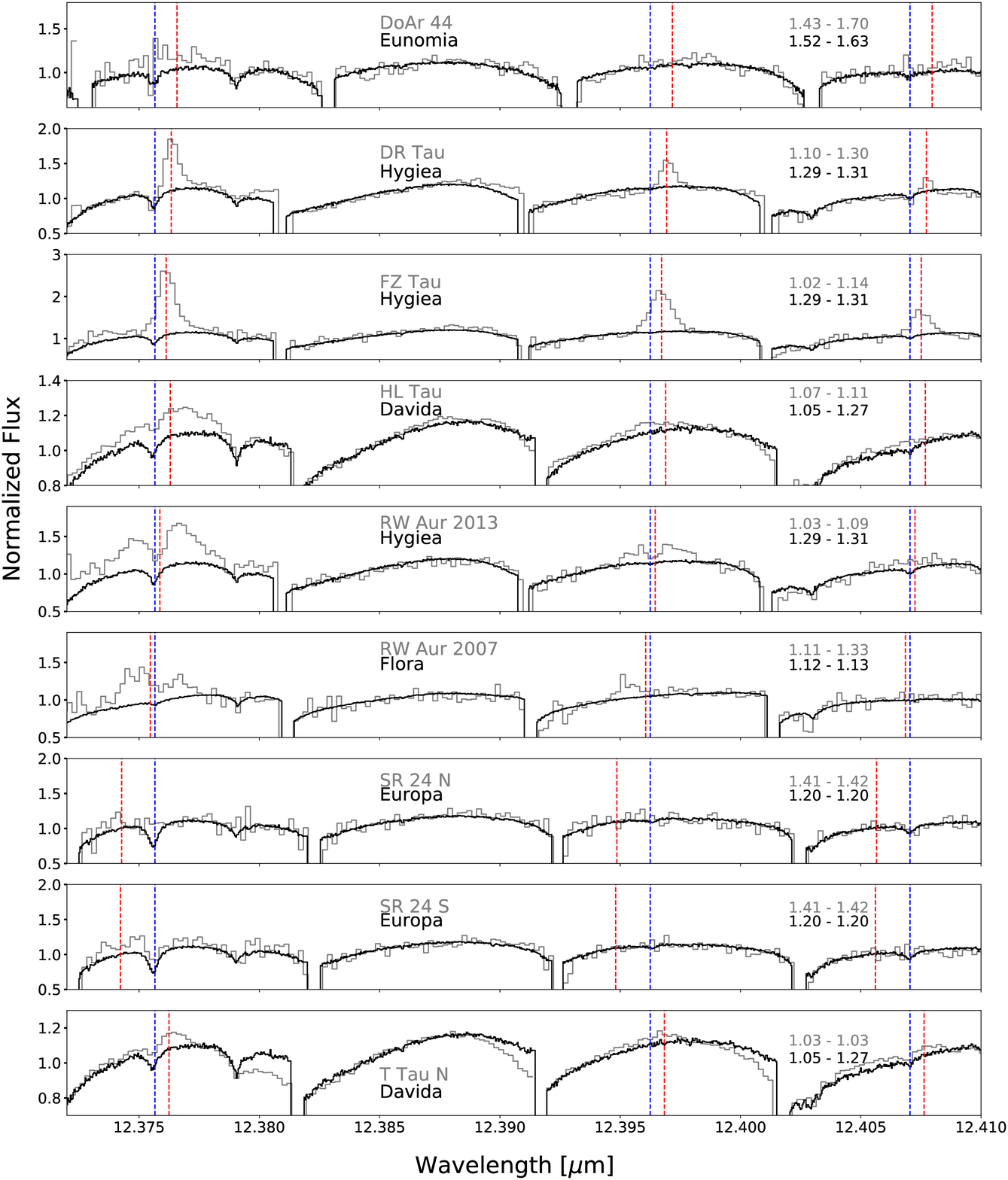}
\caption{Demonstration of telluric correction process for TEXES data.  Gray curves show binned target spectra; black curves show unbinned spectra of telluric calibrators.  Airmass ranges for the target and calibrator are labeled in gray and black, respectively.  Vertical dashed blue and red lines show the locations of telluric and source water vapor line centers.}
\label{fig:texes_telluric}
\end{figure*}

Figure \ref{fig:texes_telluric} demonstrates the telluric correction for TEXES data.   
Because the stellar water features are in some cases rather broad and weak, special care was taken in flat-fielding and correcting the spectra for telluric
features.  The stellar spectra were first flat-fielded by dividing by the
difference between spectra of a dome-temperature black body and the sky.
If the black body were at the same temperature as the atmosphere this
would correct for both instrumental effects and atmospheric absorption,
but since the black body is warmer than the atmosphere, and the atmosphere
is not isothermal, the flat-fielding procedure only corrects for about half
of the telluric absorption.  Figure \ref{fig:texes_telluric} shows both the unbinned (flat-fielded) spectra of the telluric calibrators, as well as binned (flat-felded) source spectra. Blue and red vertical lines show the expected wavelengths of telluric and source water vapor lines, respectively.

In addition, the non-linear response of the detector array and the different
illumination of the instrument pupil by the black body and the telescope result
in curvature of the echelon order spectra.  To correct for these effects, the
flat-fielded stellar spectra were divided by similarly flat-fielded spectra
of asteroids, which were adjusted for the difference in
airmass of the stars and asteroids.  This division corrected well for residual
telluric features, but in HL Tau and T Tau N left some slope or curvature
of the orders, probably because of shifts in the instrument in moving the
telescope between the star and the asteroid.  Fortunately, the residual order
shapes were quite similar among the different orders.  Therefore, after division by the standard star, the spectrum of each order was divided by a (pixel-by-pixel) median spectrum produced from the 9 orders combined.  We found that division by the median order shape improved the detectability of the emission lines from HL Tau and T Tau N, but had little effect on any of the other spectra.

\begin{figure*}
\epsscale{1.2}
\plotone{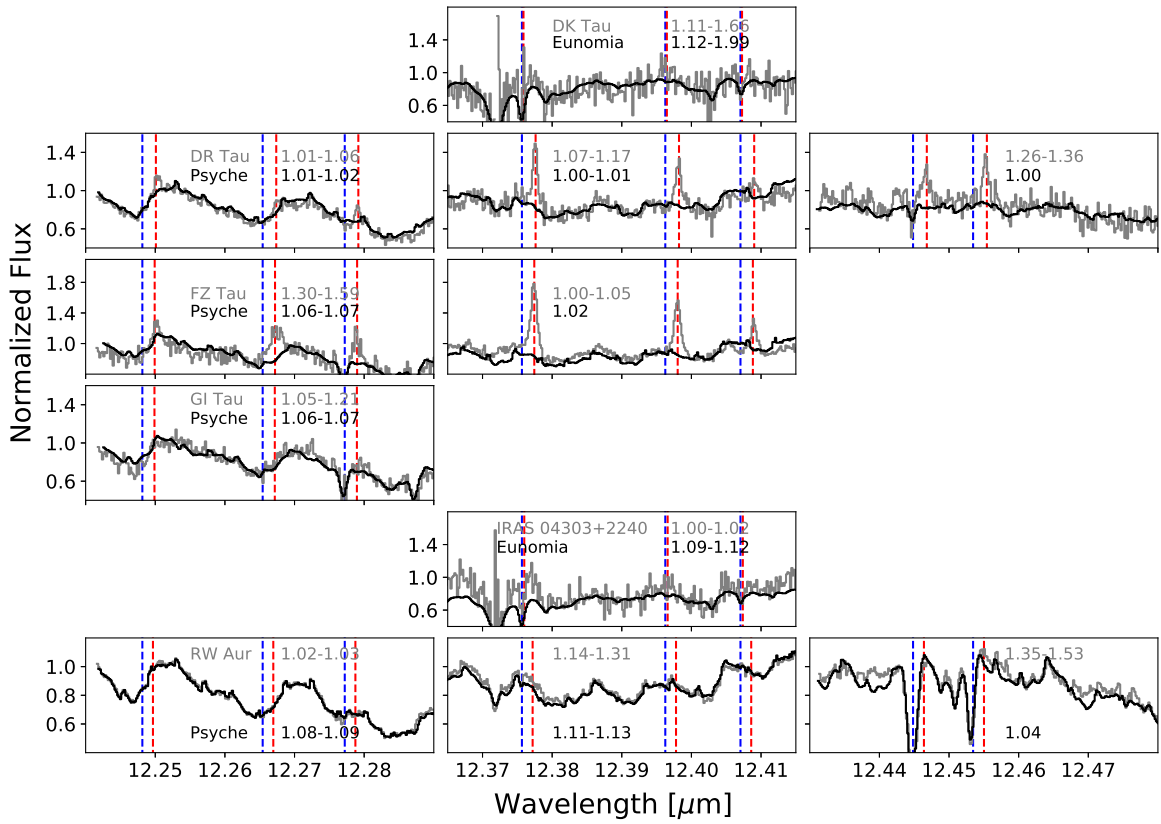}
\caption{Same as Figure \ref{fig:texes_telluric} but for Michelle spectra. Michelle spectra were not flat-fielded prior to this step, so the spectra still show significant curvature at this stage of the reduction process.}
\label{fig:michelle_telluric}
\end{figure*}

Figure \ref{fig:michelle_telluric} demonstrates the telluric correction for the Michelle data.  Flat field frames were not utilized for this dataset, as their use did not result in improvement of data quality.  Therefore, these spectra show greater order curvature and deeper atmospheric lines than in the TEXES spectra.  As in Figure \ref{fig:texes_telluric}, we show the spectra of the telluric calibrators as well as the source spectra (both unbinned), and vertical lines mark the expected wavelengths of telluric and source water vapor lines.  

\end{appendices}

\end{document}